\documentclass[a4paper,UKenglish,cleveref, autoref, cleveref, thm-restate]{lipics-v2021}

\hideLIPIcs  

\title{On Computing Vertex Connectivity of 1-Plane Graphs\footnote{A preliminary version of this paper appeared in ICALP 2023 \cite{icalp}.}} 

\author{Therese Biedl}{David R. Cheriton School of Computer Science, University of Waterloo, Canada}{biedl@uwaterloo.ca}{https://orcid.org/0000-0002-9003-3783}{}

\author{Karthik Murali}{School of Computer Science, Carleton University, Ottawa, Canada}{KarthikMurali@cmail.carleton.ca}{https://orcid.org/0000-0002-1825-0097}{}

\authorrunning{T.Biedl and K. Murali} 

\Copyright{Therese Biedl and Karthik Murali} 



\category{} 

\relatedversion{A preliminary version of this paper appeared in ICALP 2023.} 

\nolinenumbers 

\newtheorem{open_problem}{Open problem}

\newcommand{\calT}{\mathcal{T}}
\newcommand{\X}{\ensuremath{\times}}

\usepackage{framed}
\usepackage{mdframed}
\usepackage{tcolorbox}

\begin{document}

\maketitle

\abstract{
The vertex connectivity of a graph $G$ is the size of the smallest set of vertices $S$ such that $G \setminus S$ is disconnected. For the class of planar graphs, the problem of vertex connectivity is well-studied, both from structural and algorithmic perspectives. Let $G$ be a plane embedded graph, and $\Lambda(G)$ be an auxiliary graph obtained by inserting a face vertex inside each face and connecting it to all vertices of $G$ incident with the face. If $S$ is a minimal vertex cut of $G$, then there exists a cycle of length $2|S|$ whose vertices alternate between vertices of $S$ and face vertices. This structure facilitates the designing of a linear-time algorithm to find minimum vertex cuts of planar graphs. In this paper, we attempt a similar approach for the class of 1-plane graphs---these are graphs with a drawing on the plane where each edge is crossed at most once. 

We consider different classes of 1-plane graphs based on the subgraphs induced by the endpoints of crossings. For 1-plane graphs where the endpoints of every crossing induce the complete graph $K_4$, we show that the structure of minimum vertex cuts is identical to that in plane graphs, as mentioned above. For 1-plane graphs where the endpoints of every crossing induce at least three edges (i.e., one edge apart from the crossing pair of edges), we show that for any minimal vertex cut $S$, there exists a cycle of diameter $O(|S|)$ in $\Lambda(G)$ such that all vertices of $S$ are in the neighbourhood of the cycle. This structure enables us to design a linear time algorithm to compute the vertex connectivity of all such 1-plane graphs.
}


\maketitle




\section{Introduction}
 
The class of \textit{planar graphs}, which are graphs that can be drawn on the plane without crossings, is fundamental to both graph theory and graph algorithms. Many problems can be solved more efficiently in planar graphs than in general graphs. However, most real-world graphs are typically non-planar. The field of ‘beyond-planar’ graphs has recently garnered significant interest in the graph drawing community (see \cite{KLM17} for a survey and \cite{hong2020beyond} for a
book on the topic). A frequently studied class of beyond-planar graphs is \textit{1-planar graphs}---graphs that can be drawn on the plane such that each edge is crossed at most once. In this paper, we look at the problem of computing the vertex connectivity for 1-planar graphs. For any graph $G$, the \textit{vertex connectivity} of $G$, denoted $\kappa(G)$, is the size of a smallest set of vertices whose removal makes $G$ disconnected; any such set of vertices is called a \textit{minimum vertex cut} of $G$.

\subsection{Literature Review}

There is a rich and long history of results on testing vertex connectivity of general graphs. Let $G = (V,E)$ be a graph and $n := |V(G)|$ and $m := |E(G)|$. One of the earliest results is that $\kappa(G) \leq 3$ can be tested in linear (i.e. $O(m+n)$) time  \cite{tarjan1972depth, hopcroft1973dividing, DBLP:conf/gd/GutwengerM00}. For graphs with larger connectivity, Kleitman \cite{kleitman1969methods} showed how to test $\kappa(G)\leq k$ in time $O(k^2nm)$. For $\kappa(G) \in O(1)$, the first $O(n^2)$ algorithm was by Nagamochi and Ibaraki \cite{DBLP:journals/algorithmica/NagamochiI92}. The fastest deterministic algorithm is by Gabow and takes time $O(m \cdot (n + \min\{k^{5/2}, kn^{3/4}\}))$ \cite{gabow2006using}. Over the last few years, a series of randomised algorithms have brought down the run-time for testing $k$-connectivity, using state-of-art maxflow techniques, to $m^{1+o(1)}$  \cite{li2021vertex, ChenKLPGS22, nanongkai2019breaking, DBLP:conf/soda/ForsterNYSY20}. In contrast, all known deterministic algorithms are slower. Recently, Saranurak and Yingchareonthawornchai \cite{SaranurakY22} gave an $m^{1+o(1)}2^{O(k^2)}$ algorithm, which is nearly-linear for all $k \in o(\sqrt{\log n})$. The question whether there exists a linear-time deterministic algorithm for testing whether $\kappa(G) > 3$ is still open. 

\subsubsection{Planar Graphs}

For special classes of graphs such as planar graphs, vertex connectivity can be computed in linear time. Any simple planar graph $G$ has at most $3n - 6$ edges, due to which there always exists a vertex with at most five distinct neighbours---hence $\kappa(G) \leq 5$. Since $\kappa(G)\leq 3$ can be tested in linear time \cite{tarjan1972depth, hopcroft1973dividing}, it only remains to test whether $\kappa(G)=4$ or $\kappa(G) = 5$. This was solved for maximal planar graphs by Laumond \cite{laumond1990connectivity}. Later, Eppstein gave an algorithm to test vertex connectivity of all planar graphs in linear time \cite{eppstein}. 

Our results in this paper draw their inspiration from Eppstein's approach, so we briefly review it here. (To maintain consistency, we use the notation and terminology of this paper. All terms will be formally defined in \Cref{section: preliminaries}.) Let $G$ be a planar graph with a fixed planar embedding; we call such a graph a  \textit{plane graph}. Let $\Lambda(G)$ be an auxiliary plane graph obtained by adding a face vertex inside each face of $G$ and connecting it to all the vertices on the boundary of the incident face. We refer to the subgraph of $\Lambda(G)$ induced by all edges incident with face vertices as the \textit{radial graph}, and denote it by $R(G)$. For any cycle $C \subseteq R(G)$, let $\mathrm{int}_\Lambda(C)$ and $\mathrm{ext}_\Lambda(C)$ be all vertices of $\Lambda(G)$ in the interior and exterior of $C$ (but not on $C$). We call a cycle $C \subseteq R(G)$ a \textit{fence} of $\Lambda(G)$ if $\mathrm{int}_\Lambda(C) \cap V(G) \neq \emptyset$ and $\mathrm{ext}_\Lambda(C) \cap V(G) \neq \emptyset$.

In \cite{eppstein}, Eppstein shows that if $S$ is a minimal vertex cut of $G$ (i.e., no proper subset of $S$ is a vertex cut of $G$), then there exists a fence $C \subseteq R(G)$ with $V(C) \cap V(G) = S$. Conversely, for any fence $C$ of $\Lambda(G)$, the set $V(C) \cap V(G)$ is a vertex cut of $G$ as it separates $\mathrm{int}_\Lambda(C) \cap V(G)$ and $\mathrm{ext}_\Lambda(C) \cap V(G)$. Thus, computing a minimum vertex cut of a planar graph $G$ reduces to the problem of finding a shortest cycle $C \subseteq R(G)$ of length $2k$, for some $1 \leq k \leq 5$, such that $V(C) \cap V(G)$ is a vertex cut of $G$. This cycle can be found using a modification of the planar subgraph isomorphism algorithm developed in \cite{eppstein}. The algorithm runs in linear time when the size of the subgraph to be found is bounded. Having found such a cycle $C$, the set $V(C) \cap V(G)$ is returned as a minimum vertex cut of $G$. Thus, $\kappa(G)$ can be computed in linear time for all planar graphs $G$.

\subsection{Our Contribution}

With the goal of computing $\kappa(G)$ of 1-planar graphs in mind, our first contribution is to study the structure of minimal vertex cuts in 1-planar graphs. We show that, under some restrictions on the crossings in a 1-planar drawing, every minimal vertex cut of a 1-plane graph $G$ is in the neighbourhood of a fence of (a suitably defined variant of) $\Lambda(G)$. As our second contribution, we then show that---similar to Eppstein's work \cite{eppstein}---this structural result can be used to devise a linear-time algorithm to compute vertex connectivity. 

As testing 1-planarity is NP-hard even under various restrictions \cite{grigoriev2007algorithms,nphard_1planarity_parameterized,korzhik2013minimal,auer20151,cabello2013adding}, we assume our input to be a \textit{1-plane graph}---a 1-planar graph with a 1-planar embedding. 
Let $G$ be a 1-plane graph, and 
consider the plane graph $G^\times$ obtained by replacing crossing points by dummy vertices.   This defines graph $\Lambda(G^\times)$ and radial graph $R(G^\times)$; for ease of writing we will simply call these $\Lambda(G)$ and $R(G)$.
For any subgraph $H \subseteq \Lambda(G)$, let $V_\times(H) := V(H) \cap V(G^\times)$.

We consider subclasses of 1-plane graphs from the perspective of the various subgraphs induced by the endpoints of crossings (\Cref{fig: Types of crossings in a 1-plane graph}). Initially, we consider the class of \textit{full 1-plane graphs} where the endpoints of every crossing induces the complete graph $K_4$. We show that the structure of minimal vertex cuts in full 1-plane graphs is identical to that in plane graphs: for every minimal vertex cut $S$, there is a fence $C \subseteq R(G)$ such that $V_\times(C) \cap V(G) = S$. While this does not extend verbatim to 1-plane graphs that are not full, we show that a similar structure exists so long as, for every crossing, there is at least one more edge connecting the endpoints of the crossing (apart from the crossing pair of edges). We call these graphs \textit{1-plane graphs without $\times$-crossings}. (An \textit{$\times$-crossing} is one in which the endpoints induce exactly two edges, see \Cref{fig: Types of crossings in a 1-plane graph}.) For any set $Q \subseteq V(G)$, let $N_G[Q]$ be the set of all vertices of $G$ that are either in $Q$ or adjacent to some vertex of $Q$. We show that for any 1-plane graph $G$ without $\times$-crossings, and a minimal vertex cut $S$ of $G$, there exists a fence $C \subseteq R(G)$ such that $V(C) \cap V(G) \subseteq S$ and $C$ \textit{skirts} along $S$, i.e., $V_\times(C) \subseteq N_{G^\times}[S]$ and $S \subseteq N_{G^\times}[V_\times(C)]$. As 1-plane graphs are at most 7-connected \cite{Bodendiek1983BemerkungenZE,pach1997graphs}, the existence of such a cycle implies that the vertices of every minimum vertex cut are within a bounded diameter subgraph of $\Lambda(G)$. On the flip side, we show that there exist 1-plane graphs $G$ (containing many $\times$-crossings) with a unique minimum vertex cut $S$ such that no two vertices of $S$ are close to each other in $\Lambda(G)$. 

Our structural results on minimum vertex cuts also have algorithmic implications: we show that the vertex connectivity of 1-plane graphs without $\times$-crossings can be computed in linear time. To achieve this, we develop a novel concept, called a \textit{co-separating triple}, as an abstraction of the following property of plane graphs: Let $G$ be a plane graph and $C \subseteq R(G)$ be a fence. Let $A := \mathrm{int}_\Lambda(C)$, $X := V(C)$, and $B := \mathrm{ext}_\Lambda(C)$. Notice that $X$ separates $A$ and $B$ in $\Lambda(G)$, while simultaneously, $X \cap V(G)$ separates $A \cap V(G)$ and $B \cap V(G)$ in $G$. We refer to any three sets $(A,X,B)$ that satisfy this property as a \textit{co-separating triple}. In the above example of plane graphs, if $C \subseteq R(G)$ is the fence of length $2|S|$ that contains all vertices of a minimum vertex cut $S$, then $N_G[X \cap V(G)]$ has bounded diameter in $\Lambda(G)$, since $X \cap V(G) = S$ and $S \in O(1)$. 

Let $S$ be a minimum vertex cut of $G$, and recall that there exists a fence $C \subseteq R(G)$ that skirts along $S$ such that $V(C) \cap V(G) \subseteq S$. We let $X := V(C) \cup S$, and $A,B$ be all the remaining vertices in $\mathrm{int}_\Lambda(C)$ and $\mathrm{ext}_\Lambda(C)$, respectively. We show that $(A,X,B)$ is a co-separating triple. Since vertices of $X$ are in the neighbourhood of $S$, and connected via $C$, we can argue that any two vertices of $X$ are within distance $O(|S|) \subseteq O(1)$ in $\Lambda(G)$. This helps us restrict our search for a co-separating triple among various subgraphs of $\Lambda(G)$ with bounded diameter. As $\Lambda(G)$ is planar, and planar graphs with small diameter have small treewidth, we can use the framework of Monadic Second Order Logic (MSOL) and Courcelle's theorem to find a co-separating triple, and return a minimum vertex cut in linear time.

\subparagraph{Follow-up Work.}
The current paper makes three major contributions: (1) Show that for every minimal vertex cut $S$ in a 1-plane graph without $\times$-crossings, there is a fence that skirts along $S$ (in addition, if the graph is full 1-plane, the $G$-vertices on the fence are precisely the set of $S$-vertices); (2) show that the fence induces a co-separating triple $(A,X,B)$ where all vertices of $X$ are close in $\Lambda(G)$; and (3) develop a linear-time algorithm to find such a co-separating triple and obtain a minimum vertex cut. In a follow-up paper \cite{esa_paper}, we focus on understanding structural conditions related to crossings that ensure every minimal vertex cut gives rise to a co-separating triple where vertices of $X$ are close in $\Lambda(G)$. This implies result (3) in the same way, and permits us to compute vertex connectivity for a wider class of near-planar graphs such as $k$-plane graphs with bounded number of $\times$-crossings, $d$-map graphs, optimal $k$-planar graphs for $k\in \{1,2,3\}$, graphs with small crossing number, $d$-framed graphs, etc. From an algorithmic perspective, the follow-up paper \cite{esa_paper} fundamentally relies on the theoretical groundwork laid here. Although the results in \cite{esa_paper} subsume those presented here, the primary objective of this paper is to demonstrate how the well-known equivalence between minimal vertex cuts and fences in planar graphs extends to the 1-plane setting. Extending this equivalence to the broader classes of graphs considered in \cite{esa_paper} may not be possible.


\subparagraph{Organisation of the Paper.} In \Cref{section: preliminaries}, we give all definitions that are preliminary to the paper. Then, in \Cref{sec: building blocks}, we provide a more detailed review of the graph $\Lambda(G)$ and the concept of a co-separating triple. After that, as a first step, we look at the class of full 1-plane graphs in \Cref{section: Vertex connectivity of full 1-plane graphs}. Indeed, we use the results therein when we consider 1-plane graphs without $\times$-crossings in \Cref{section: vertex connectivity of general 1-plane graphs}. \Cref{section: Vertex connectivity of full 1-plane graphs,section: vertex connectivity of general 1-plane graphs} are devoted to understanding the structure of minimal vertex cuts. Using these structural results, we then design a linear-time algorithm for testing vertex connectivity in \Cref{section: computing vertex connectivity in linear time}. We conclude the paper in \Cref{sec: conclusion} with a few open questions. 


\section{Preliminaries}\label{section: preliminaries}

\begin{figure}
    \centering
    \includegraphics[width=\linewidth,page=2]{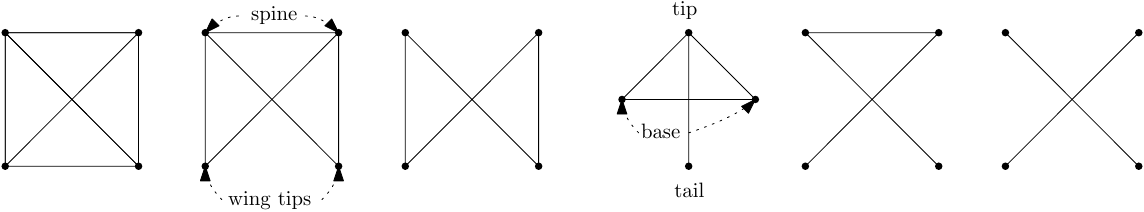}
    \caption{Crossings in 1-plane graphs: full, almost-full, bowtie, arrow, chair and \X.
}
\label{fig: Types of crossings in a 1-plane graph}
\end{figure}

All graphs in this paper are connected and without loops, but we permit them to have parallel edges since these do not affect connectivity, but will be helpful when making other assumptions later.
A \textit{vertex cut} of a graph $G$ is a set $S$ of vertices such that $G\setminus S$ is disconnected; we use the term \textit{flap} for a connected component of $G\setminus S$. A set $S$ is a \textit{minimal vertex cut} of $G$ if no proper subset of $S$ is a vertex cut of $G$. A set $S$ is a minimal vertex cut if and only if every vertex of $S$ has a neighbour in each flap of $G \setminus S$. A \textit{minimum vertex cut} of $G$ is a vertex cut $S$ such that $|S|$ is the minimum possible; $\kappa(G) := |S|$ is called the \emph{vertex connectivity} of $G$. A set $S$ \textit{separates} two non-empty sets $A,B$ in a graph $G$ if no two vertices $a \in A$ and $b \in B$ belong to the same flap of $G \setminus S$. Equivalently, every $(a,b)$-path (a path with $\{a,b\}$ as end vertices) contains a vertex of $S$. The length of a path $P$ is the number of edges in $P$, and will be denoted by $|P|$. 
For any set of vertices $Q$ in a graph $G$, we use the notation $N_G(Q)$ to denote the set of all vertices of $G$ that are adjacent to some vertex of $Q$; we use $N_G[Q]$ to denote the union $Q \cup N_G(Q)$. For any set $Z \subseteq V(G)$, we write \textit{$Z$-vertex} as a shortcut for vertex of $Z$.

\subparagraph{Planar and 1-Planar Graphs.} A \textit{drawing} of a graph $G = (V,E)$ is a mapping of each vertex of $V(G)$ to a distinct point on the plane, and each edge $(u,v)$ of $E(G)$ to a simple 
curve that connects the points representing $u$ and $v$ and that does not pass through any other vertices. 
Further, we may assume that no pair of edges touch each other tangentially, and no three edges intersect at a point interior to all three edges. A pair of edges $\{e_1,e_2\}$ \textit{cross} in the drawing if there is a point in the plane that is common to the interior of both curves representing $e_1$ and $e_2$. A graph $G$ is a \textit{planar graph} if it can be drawn on the plane without crossings; a graph $G$ with such a drawing is called a \textit{plane graph}. A \textit{face} of a plane graph is a maximal region of the plane that does not intersect the drawing. The boundaries of the faces of a connected plane graph can be uniquely determined 
by its \textit{rotation system}, which is the set of \textit{rotations} $\rho(v)$ at each vertex $v$, where a rotation is the clockwise sequence of edges incident with $v$. 

A \textit{1-planar graph} is a graph that has a drawing in the plane where each edge is crossed at most once. 
A graph with such a drawing is called a \textit{1-plane graph}. While one can decide in linear time whether a graph $G$ is planar or not \cite{hopcroft1974efficient}, it is NP-hard to decide whether a graph is 1-planar \cite{grigoriev2007algorithms,nphard_1planarity_parameterized,korzhik2013minimal,auer20151,cabello2013adding}. Therefore, we assume that all 1-planar graphs come with a 1-planar drawing; in other words, they are 1-plane graphs. We may assume that all 1-plane graphs are drawn such that no crossing pair of edges are incident with the same vertex; this is not a restriction, as one can re-draw edges with local changes to avoid such crossings. The \textit{planarization} of a 1-plane graph $G$, denoted $G^\times$, is the plane graph obtained by replacing any point where edges cross with a \textit{dummy vertex}. For any subdrawing $H$ of $G$, we let $H^\times$ be the subgraph of $G^\times$ obtained by replacing crossing points on edges of $H$ with dummy vertices. For instance, if $e = (u,v)$ is an edge in $E(G)$, then $e^\times = e$ if $e$ is uncrossed, else $e^\times$ is a path $(u,c,v)$ where $c$ is the crossing point on $(u,v)$. The \textit{faces} of a 1-plane graph are the faces of $G^\times$, and these can be uniquely determined by the rotation system of $G^\times$. 

\subparagraph{Crossings in 1-Plane Graphs.} Let $\{(u,v),(w,x)\}$ be a crossing in a 1-plane graph $G$, i.e., edges $(u,v)$ and $(w,x)$ cross each other. The four vertices $\{u,v,w,x\}$ are called \emph{endpoints of the crossing}. Two endpoints of the crossing are called \emph{consecutive} if they are not incident to the same edge; else they are called \textit{opposite}. One can classify crossings into six types based on the subgraphs induced by the four endpoints (up to considering parallel edges as a single edge). Refer to Figure~\ref{fig: Types of crossings in a 1-plane graph} for an illustration. A crossing $\{(u,v), (w,x)\}$ is called \emph{full} if its endpoints induce the complete graph $K_4$, and \emph{almost-full} if they induce $K_4$ minus one edge (the terms full and almost-full are borrowed from \cite{FHM+20}). The crossing is called \emph{bowtie} if its endpoints induce a cycle, \emph{arrow} if they induce $K_{1,3}$ plus one edge, \emph{chair} if they induce a path on three edges, and $\times$-crossing if they induce a matching. 

For later sections of the paper, we will need to distinguish the four endpoints of almost-full crossings and arrow crossings. The \textit{wing tips} of an almost-full crossing are the pair of non-adjacent endpoints, and the remaining two vertices are called \emph{spine vertices}. The \textit{tail} of an arrow crossing is the vertex not adjacent to any consecutive endpoint, and the \textit{tip} is the endpoint opposite to the tail. The \textit{base vertices} of an arrow crossing are the two endpoints apart from the tail and tip.

Let $\{(u,v), (w,x)\}$ be a pair of edges crossing at a point $c$, and $(u,x) \in E(G)$. If the edges $\{(c, u), (u,x), (x,c)\}$ bound a face in $G^\times$, then $(u,x)$ is called a \textit{kite edge} of the crossing, and the face is called a \textit{kite face} of the crossing. In a 1-planar drawing, it is not necessary that edges connecting consecutive endpoints of the crossing be kite edges. For purposes of testing vertex connectivity, if an edge such as $(u,x)$ in the above crossing is not a kite edge, we can duplicate it and insert it as a kite edge by simply drawing it close to the curve $u\textrm{-}c\textrm{-}x$ (adding parallel edges does not affect the vertex connectivity of a graph). 
Henceforth, we may assume that at any crossing, all edges between consecutive endpoints are kite edges.


\section{Building Blocks}\label{sec: building blocks}

This section introduces two new concepts---radial planarisation and co-separating triple.
Both these concepts, foundational to this paper, have been inspired from ideas on testing connectivity for planar graphs \cite{eppstein}. These concepts are also the building blocks for the sequel paper \cite{esa_paper} on testing vertex and edge connectivity for a wider class of near-planar graphs.    

\subsection{Radial Planarisation}

The first step in computing the vertex connectivity of a plane graph is the following \cite{eppstein}: Construct an auxiliary plane graph by adding a face vertex inside each face of $G$, and make it adjacent to all vertices on the face boundary. We generalise this construction to 1-plane graphs by repeating the same process on the planarisation $G^\times$---the resulting graph is called the radial planarisation of $G$.

 \begin{figure}
   \begin{subfigure}[b]{0.45\textwidth}
     \includegraphics[page=2,width=\linewidth]{Figures/Radial_graph_counter_example.pdf}
   \end{subfigure}
   \hspace*{\fill}
   \begin{subfigure}[b]{0.45\textwidth}
     \centering
     \includegraphics[page=1,width=\linewidth]{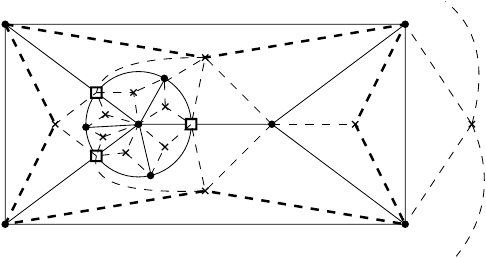}
   \end{subfigure}
   \caption{An example to illustrate radial planarisation. On the left is a 1-plane graph $G$ with three arrow crossings. On the right, the radial graph $R(G)$ is indicated by dashed edges. The dummy vertices are indicated by hollow squares. The bold-dashed edges form a cycle $C$ that is not a fence as $\mathrm{ext}_G(C) = \emptyset$.}
   \label{fig: Radialisation}
 \end{figure}

\begin{definition}[Radial Planarisation]
For a 1-plane graph $G$, the \textit{radial planarization} of $G$, denoted $\Lambda(G)$, is obtained as follows: Insert a \textit{face vertex} $f$ inside each face $F$ of $G^\times$, and for every incidence of $F$ with a vertex $v$, add an edge $(v,f)$ inside $F$. 
\end{definition}

(Refer to Figure \ref{fig: Radialisation} for an illustration.) The graph $\Lambda(G)$ has three types of vertices: $G$-vertices (vertices of $G$), dummy vertices (at crossing points of $G$), and face vertices. For any subgraph $H \subseteq \Lambda(G)$, we use the notation $V_\times(H)$ to denote the set $V(H) \cap V(G^\times)$, i.e., all $G$-vertices and dummy vertices on $H$. We call the plane bipartite subgraph of $\Lambda(G)$ induced by edges incident with face vertices the \textit{radial graph}, and denote it by $R(G)$. Note that $R(G)$ need not be a simple graph, since multiple incidences of a vertex with a face give rise to parallel edges. Although $R(G)$ may not be simple, it does not contain \textit{bigons}---faces bound by exactly two edges. 

For any plane graph $G$, the vertices of any minimal vertex cut $S$ can be connected by a cycle $C \subseteq R(G)$ such that $V(C) \cap V(G) = S$, and there exist $G$-vertices inside and outside the cycle \cite{eppstein}. In this paper, we will see that cycles of this type also occur in 1-plane graphs without $\times$-crossings. For convenience, we develop some notations associated with such cycles. For any cycle $C \subseteq R(G)$, let $\mathrm{int}_\Lambda(C)$ and $\mathrm{ext}_\Lambda(C)$ denote the set of all vertices of $\Lambda(G)$ that are in the interior and exterior of $C$ (but not on $C$). Let $\mathrm{int}_G(C) = \mathrm{int}_\Lambda(C) \cap V(G)$ and $\mathrm{ext}_G(C) = \mathrm{ext}_\Lambda(C) \cap V(G)$. We say that a cycle $C \subseteq R(G)$ is a \textit{fence} of $\Lambda(G)$ if $\mathrm{int}_G(C) \neq \emptyset$ and $\mathrm{ext}_G(C) \neq \emptyset$. For any set of vertices $Q \subseteq V(G)$, we say that a fence $C$ \textit{skirts} along $Q$ if $Q \subseteq N_{G^\times}[V(C)]$ and $V(C) \subseteq N_{G^\times}[Q]$. 
With this notation, one can now re-phrase the above-mentioned property about plane graphs as follows: 

\begin{theorem}[Lemma 8 in \cite{eppstein}]\label{theorem: vertex connectivity planar graphs}
\label{thm:planar}
Let $G$ be a plane graph and $S$ be a minimal vertex cut of $G$. Then there is a fence $C \subseteq R(G)$ such that $V(C) \cap V(G)=S$.
\end{theorem}

Notice that in a plane graph $G$, if $C$ is a fence of $\Lambda(G)$, then $V(C) \cap V(G)$ is a vertex cut of $G$, and $V(C)$ is a vertex cut of $\Lambda(G)$. This, together with \Cref{theorem: vertex connectivity planar graphs}, implies that the problem of computing vertex connectivity reduces to the problem of finding a fence of the smallest length in $\Lambda(G)$. Since a planar graph is at most 5-connected, there exists a fence of length at most 10 in $\Lambda(G)$. Using this observation, Eppstein \cite{eppstein} then uses his planar subgraph isomorphism algorithm, which runs in linear time for subgraphs of small diameter, to find a shortest cycle $C \subseteq R(G)$ that separates two $G$-vertices in $\Lambda(G)$\footnote{In Theorem 8 of \cite{eppstein}, Eppstein writes that it is sufficient to find $C \subseteq R(G)$ that separates two $G$-vertices in $R(G)$---but this is incorrect. In Figure \ref{fig: Radialisation}, the cycle indicated by the bold dashed edges separates $s$ and $t$ in $R(G)$, but not in $G$.}. 

\subsection{Co-Separating Triple}

In the description above, the crucial insight was the existence of a vertex set $X \subseteq V(\Lambda(G))$ of small diameter in $\Lambda(G)$ such that $X$ is a vertex cut of $\Lambda(G)$ and $X \cap V(G)$ is a vertex cut of $G$. We call such a structure a co-separating triple, and express the diameter property through a sub-structure called the nucleus. Since we will use co-separating triples not only in the context of $\Lambda(G)$, but for various related graphs derived from it (see \Cref{section: computing vertex connectivity in linear time}), we give a generic definition in terms of an abstract graph $\Lambda$ where $V(\Lambda) \subseteq V(\Lambda(G))$. (For \Cref{def: co-separating triple}, recall the notation $e^\times$ for the path in $G^\times$ corresponding to an edge $e \in E(G)$.)

\begin{restatable}[Co-separating Triple]{mydef}{CoSeparatingTriple}
\label{def: co-separating triple}
Let $G$ be a 1-plane graph, and $\Lambda$ be a graph such that $V(\Lambda) \subseteq V(\Lambda(G))$. A partition of $V(\Lambda)$ into three sets $(A,X,B)$ is called a \emph{co-separating triple of $(\Lambda, G)$} if it satisfies the following properties:

\begin{enumerate}
\item Each of $A$, $X$ and $B$ contains at least one vertex of $G$.\label{c1}
\item There is no edge of $\Lambda$ with one end in $A$ and the other in $B$.\label{c2}
\item For every edge $e \in G \setminus (X \cap V(G))$, all vertices of $e^\times \cap \Lambda$ belong entirely to $A \cup X$, or to $B \cup X$.\label{c3}
\end{enumerate}
\end{restatable}

\medskip
The name ``co-separating triple'' is motivated by the observation that $X$ simultaneously  gives vertex cuts in $G$ and $\Lambda(G)$, as proved below.

\begin{observation}\label{obs: co-sep triple}
If $(A,X,B)$ is a co-separating triple of $(\Lambda,G)$, then $X$ separates $A$ and $B$ in $\Lambda$ while $X \cap V(G)$ separates $A \cap V(G)$ and $B \cap V(G)$ in $G$.
\end{observation}

\begin{proof}
By Condition \ref{c2} of \Cref{def: co-separating triple}, there is no edge in $\Lambda$ between $A$ and $B$. By Condition \ref{c3}, there is no edge of $G$ with one end in $A$ and the other in $B$. Condition \ref{c1} ensures that $A$ and $B$ are non-empty. Therefore, $X$ separates $A$ and $B$ in $\Lambda$, and $X \cap V(G)$ separates $A \cap V(G)$ and $B \cap V(G)$ in $G$. 
\end{proof}

To compute co-separating triples later, we will explicitly search for vertex set $X$ in a subgraph of $\Lambda(G)$. However, we will not compute all of $A$ and $B$, and instead only search for a subset of their vertices to guarantee their existence.   To describe this, the following notation will be helpful.

\begin{definition}[Nucleus of a Co-separating Triple]
The nucleus of a co-separating triple $(A,X,B)$ of $(\Lambda,G)$ is the set $N_G[X \cap V(G)]$. The maximum distance in $\Lambda$ between any two vertices of the nucleus is called the \textit{nuclear diameter}.   
\end{definition}

\begin{observation}\label{obs: nucleus elements}
The nucleus of any co-separating triple $(A,X,B)$ contains a vertex of $A \cap V(G)$ and a vertex of $B \cap V(G)$.
\end{observation}
\begin{proof}
Let $a \in A \cap V(G)$ and $b \in B \cap V(G)$ be two vertices (these exist by Condition \ref{c1} of \Cref{def: co-separating triple}). Consider any $(a,b)$-path in $G$. As $X \cap V(G)$ separates $A \cap V(G)$ from $B \cap V(G)$ (\Cref{obs: co-sep triple}), this path contains vertices $a' \in A \cap V(G)$ and $b' \in B \cap V(G)$ such that $\{a',b'\} \subseteq N_G[X \cap V(G)]$.
\end{proof}

Any fence $C$ of a plane graph $G$ provides a simple way to define a co-separating triple $(A,X,B)$ of $(\Lambda(G),G)$---simply choose $A := \mathrm{int}_\Lambda(C)$, $X := V(C)$ and $B := \mathrm{ext}_\Lambda(C)$ (\Cref{thm: co-separating triple plane graph} shows why this works). As a matter of fact, all co-separating triples in this paper will be obtained by similar means---we first show (as in \Cref{theorem: vertex connectivity planar graphs}) that for any minimal vertex cut $S$ in a 1-plane graph without $\times$-crossings, there exists of a fence $C \subseteq R(G)$ that, roughly speaking, connects vertices of $S$, and $(A,X,B)$ are then the vertices inside, on, and outside $C$.

\begin{theorem}\label{thm: co-separating triple plane graph}
    Let $G$ be a plane graph and $S$ be a minimal vertex cut of $G$. Then there exists a co-separating triple $(A,X,B)$ of $(\Lambda(G),G)$ such that $X \cap V(G) = S$ and the nuclear diameter is at most $|S|+2$.
\end{theorem}

\begin{proof}
From \Cref{theorem: vertex connectivity planar graphs}, there exists a fence $C \subseteq R(G)$ such that $V(C) \cap V(G) = S$. Let $A = \mathrm{int}_\Lambda(C)$, $X = V(C)$ and $B = \mathrm{ext}_\Lambda(C)$. The sets $(A,X,B)$ clearly partition $V(\Lambda(G))$, and $X \cap V(G) = S$. Both $A$ and $B$ contain vertices of $G$ because $C$ is a fence. As $C$ is a closed Jordan curve, there is no edge of $E(G) \cup E(\Lambda(G))$ with one endpoint in $A$ and the other endpoint in $B$. Therefore, $(A,X,B)$ is a co-separating triple. It is easy to see that the nucleus $N_G[X \cap V(G)] = N_G[S]$ has diameter at most $|S|+2$. For any pair of vertices $u,v \in S$, one can find a path along $C$ of length at most $|S|$, since $C$ has length $2|S|$. Therefore, for any pair of vertices in $N_G[S]$, there is a path of length at most $|S| + 2$.  
\end{proof}
  

\section{Full 1-Plane Graphs}\label{section: Vertex connectivity of full 1-plane graphs}
\label{SEC:FULL}

Now, we show that one can extend Theorem \ref{theorem: vertex connectivity planar graphs} to \textit{full 1-plane graphs}: 1-plane graphs where the endpoints of every crossing induces a $K_4$. (For Theorem \ref{theorem: vertex connectivity full 1-plane graph}, recall the notation $V_\times(H)$ to denote all vertices of $H$ that are in $V(G^\times)$, i.e., 
$G$-vertices and dummy vertices on $H$.) 

\begin{theorem}\label{theorem: vertex connectivity full 1-plane graph}
\label{thm:full}
Let $G$ be a full 1-plane graph and $S$ be a minimal vertex cut of $G$. Then there is a fence $C \subseteq R(G)$ such that $V_\times(C) = S$; in particular, $C$ contains no dummy-vertices.
\end{theorem}

\begin{proof}
We first give an outline. To construct the desired fence, we make use of face vertices of some special faces called transition faces. Let a face $F$ of $G^\times$ be a \emph{transition face} if $F$ is either incident to an edge between two $S$-vertices, or $F$ is incident to vertices from different flaps of $G\setminus S$; the corresponding face vertex in $\Lambda(G)$ is called a \emph{transition-face vertex}. Our main idea for proving Theorem \ref{theorem: vertex connectivity full 1-plane graph} is to show that every transition-face vertex is incident to two edges connecting it to $S$-vertices (\Cref{claim: marked face vertex degree 2}), and conversely, every $S$-vertex is incident to two edges connecting it to transition-face vertices (\Cref{claim: full 1-plane case analysis}). With this, one can construct a maximal path in $R(G)$ that alternates between $S$-vertices and transition-face vertices, and then close it up to form a fence $C$ such that $V_\times(C) = S$. Below, we expand upon the details formally. 

\begin{lemma}\label{claim: marked face vertex degree 2}
Every transition-face vertex is incident with at least two edges connecting it to $S$-vertices.
\end{lemma}

\begin{proof}
Let $F$ be a transition face. The statement clearly holds if $F$ is incident to an edge between two $S$-vertices; hence assume that $F$ contains vertices $u\in \phi_1$ and $w\in \phi_2$, for some pair of flaps $\phi_1,\phi_2$ of $G\setminus S$. Consider a closed walk along the boundary of $F$ from $u$ to $w$ and back to $u$. While walking from $u$ to $w$, the walk transitions from vertices in $\phi_1$ to vertices in $\phi_2$. This transition cannot happen at a dummy vertex because $G$ is full 1-plane:
(any face incident to a dummy vertex is a kite face; the incident kite edge cannot have its endpoints in both $\phi_1$ and $\phi_2$). 
Therefore, the transition from $\phi_1$ to $\phi_2$ must happen at an $S$-vertex. This accounts for one edge incident with the face vertex of $F$ with an $S$-vertex. A second such edge can be found when we transition from $\phi_2$ to $\phi_1$ while walking from $w$ to $u$ along $F$.
\end{proof}

In Lemma \ref{claim: full 1-plane case analysis}, we show that every $S$-vertex is in turn incident with at least two edges that connect it to transition-face vertices.

 \begin{figure}
   \begin{subfigure}[b]{0.3\textwidth}
     \includegraphics[page=1,width=\linewidth]{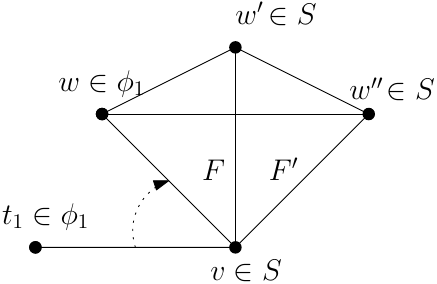}
     \caption{Case (a)}
   \end{subfigure}
   \hspace*{\fill}
   \begin{subfigure}[b]{0.3\textwidth}
     \centering
     \includegraphics[page=2,width=\linewidth]{Figures/FindTransitionFace.pdf}
     \caption{Case (b)}
   \end{subfigure}
   \hspace*{\fill}
   \begin{subfigure}[b]{0.3\textwidth}
     \centering
     \includegraphics[page=3,width=\linewidth]{Figures/FindTransitionFace.pdf}
     \caption{Case (c)}
   \end{subfigure}
   \caption{Finding a transition-face between $(v,t_1)$ and $(v,t_2)$.}
   \label{fig:FindTransitionFace}
 \end{figure}

\begin{lemma}\label{claim: full 1-plane case analysis}
Let $(v,t_1)$ and $(v,t_2)$ be two edges of $G$, where $v \in S$
and $t_1,t_2$ belong to different flaps of $G\setminus S$. Let $\rho(v)$ be the clockwise sequence of edges of $G \cup R(G)$ incident to $v$. Then there exist transition-face vertices $f_1$ and $f_2$ adjacent to $v$ such that $(v,t_1), (v,f_1), (v,t_2), (v,f_2)$
occur in this order in $\rho(v)$.
\end{lemma}

\begin{proof}
We will show the existence of edge $(v,f_1)$ between $(v,t_1)$ and $(v,t_2)$, and a symmetric argument shows that $(v,f_2)$ between $(v,t_2)$ and $(v,t_1)$. Consider the sequence of edges in $\rho(v)$ from $(v,t_1)$ until $(v,t_2)$. Let $(v,w)$ be the last edge for which $w$ is in the same flap as $t_1$, say flap $\phi_1$. Let the next few entries in $\rho(v)$ be $(v,w),(v,f),(v,w'),(v,f'),(v,w'')$; let $F,F'$ be the faces corresponding to $f,f'$. We consider three cases, and in each of them, we show the that an edge $(v,f_1)$ exists as required (see Figure \ref{fig:FindTransitionFace} for illustrations).

\smallskip
\noindent \textbf{Case (a): $(v,w')$ is crossed.} Since $G$ is full 1-plane, this crossing has four kite faces, and since these occur before/after the crossed edge in the rotation, $\{F,F'\}$ must be kite faces of this crossing (Figure \ref{fig:FindTransitionFace}(a)). Hence the edge that crosses $(v,w')$ is $(w,w'')$, and we have kite edges $(w,w')$ and $(w',w'')$. As $w'$ and $w''$ are adjacent to $w$, and by our choice of vertex $w$, we must have $\{w',w''\} \subseteq S$. Hence $F'$ is incident to edge $(v,w'')$ with both endpoints in $S$, by which $f'$ is a transition-face vertex.

\smallskip
\noindent \textbf{Case (b): $(v,w)$ is crossed.} This makes $F$ a kite face with $(w,w')$ as a kite edge (Figure \ref{fig:FindTransitionFace}(b)). By arguments similar to Case (a), $w'\in S$, making $f$ a transition-face vertex.

\smallskip
\noindent \textbf{Case (c): Both $(v,w')$ and $(v,w)$ are uncrossed.} If $w'\in S$, then $F$ is incident to edge $(v,w')$ with both endpoints in $S$ (Figure \ref{fig:FindTransitionFace}(c)). If $w'\not\in S$, then by $w'\not\in \phi_1$, vertex $w'$ belongs to a different flap than $\phi_1$. Either way, $f$ is a transition-face vertex.
\end{proof}

Using Lemmas \ref{claim: marked face vertex degree 2} and \ref{claim: full 1-plane case analysis}, we can construct a fence $C$ such that $V_\times(C) = S$. 
Starting with an arbitrary $S$-vertex $v_1$, construct a simple path $P = v_1f_1v_2\dots f_{k-1}v_k$ alternating between transition-face vertices and $S$-vertices that is maximal in the following sense: $v_k \in S$, and for any transition-face vertex $f_{k}$ and any $S$-vertex $v_{k+1}$, the extension $P \cup \{(v_k,f_{k}),(f_{k},v_{k+1})\}$ is not a simple path. 
Therefore, for any transition-face vertex $f_{k}$ adjacent to $v_k$, and any $S$-vertex $v_{k+1}$ adjacent to $v_{k+1}$, at least one of $\{f_{k},v_{k+1}\}$ belongs to $P$. 
Since $v_k \in S$, and $S$ is minimal, $v_k$ has neighbours $t_1,t_2$ in different flaps of $G\setminus S$. By Lemma \ref{claim: full 1-plane case analysis}, there exist transition-face vertices $f'$ and $f''$ such that $(v_k,t_1), (v_k,f'), (v_k,t_2), (v_k,f'')$ occur in this order in $\rho(v)$. If $k>1$, then (up to renaming of $t_1,t_2$) we may assume that $f_{k-1} = f''$. Hence, we may assume that the edge $(v_k,f')$ is not in $P$. We now consider cases:

\smallskip
\noindent \textbf{Case (a): $f' \in P$.} Say $f' = f_i$ for some $1 \leq i \leq k-1$ 
(Figure \ref{fig:Separating cycles}(a)). Then $C:= f_iv_{i+1}\dots v_kf_i$ is a fence with $t_1$ and $t_2$ on different sides.

\smallskip
\noindent \textbf{Case (b): $f' \notin P$.} By Lemma \ref{claim: marked face vertex degree 2}, $f'$ is incident with two edges $e,e'$ that connect it to $S$-vertices. Up to renaming, we may assume that $e=(v_k,f')$. Consider extending $P$ via $e$ and $e'$. Then $P \cup \{e,e'\}$ is non-simple (as $P$ is maximal), and by $f'\notin P$, we have $e'=(f',v_i)$, for some $1 \leq i \leq k$. If $1 \leq i \leq k-1$ (Figure \ref{fig:Separating cycles}(b)), then define simple cycle $C := v_if_{i+1}\dots v_kf'v_i$ and observe that it has $t_1$ and $t_2$ on different sides; therefore $C$ is a fence. Otherwise, if $i=k$, both edges $e,e'$ connect $v_k$ to $f'$ (Figure \ref{fig:Separating cycles}(c)), and we let $C$ be the cycle consisting of $e$ and $e'$. As $R(G)$ contains no bigons, $C$ must be a fence.

\medskip
In both cases, we have constructed a fence $C \subseteq R(G)$ with
$V_\times(C) \subseteq S$. To show that $V_\times(C) = S$, it is sufficient (by minimality of $S$) to show that $V_\times(C)$ is a vertex cut of $G$. To see this, fix any path
$P$ from $a$ to $b$ in $G$, where $a \in \mathrm{int}_G(C)$ and $b \in \mathrm{ext}_G(C)$. The path $P$ must intersect $C$, by planarity of $G^\times$, and it must do so at an $S$-vertex, as $C$ contains no dummy vertices. Therefore, $V_\times(C)$ is a vertex cut of $G$, showing that $V_\times(C) = S$.
\end{proof}

 \begin{figure}
   \begin{subfigure}[b]{0.32\textwidth}
     \includegraphics[width=\linewidth]{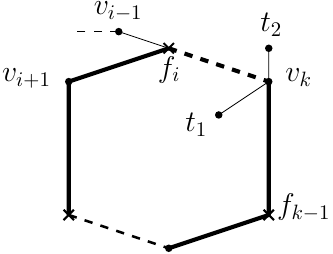}
      \caption{ }
     \label{subfig: Separating cycles case 1}
   \end{subfigure}
   \hspace*{\fill}
   \begin{subfigure}[b]{0.32\textwidth}
     \centering
     \includegraphics[width=\linewidth]{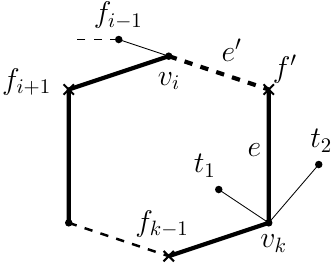}
     \caption{ }
     \label{subfig: Separating cycles case 2}
   \end{subfigure}
   \hspace*{\fill}
   \begin{subfigure}[b]{0.14\textwidth}
     \centering
     \includegraphics[width=\linewidth]{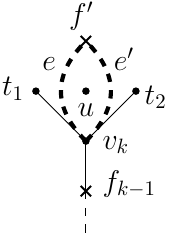}
     \caption{ }
     \label{subfig: Separating cycles case 3}
   \end{subfigure}
   \caption{Constructing a fence in a full 1-plane graph.}
   \label{fig: Constructing a marked separating cycle}
\label{fig:Separating cycles}
 \end{figure}

Similar to \Cref{thm: co-separating triple plane graph}, we show that the existence of a fence as in \Cref{theorem: vertex connectivity full 1-plane graph} gives rise to a co-separating triple of $(\Lambda(G),G)$ with small nuclear diameter. 

\begin{theorem}\label{thm: co-separating triple full 1-plane}
    Let $G$ be a full 1-plane graph and $S$ be a minimal vertex cut of $G$. Then there exists a co-separating triple $(A,X,B)$ of $(\Lambda(G),G)$ such that $X \cap V(G) = S$ and the nuclear diameter is at most $|S|+4$.
\end{theorem}

\begin{proof}
From \Cref{theorem: vertex connectivity full 1-plane graph}, there exists a fence $C$ such that $V_\times(C) = S$. Let $A = \mathrm{int}_\Lambda(C)$, $X = V(C)$ and $B = \mathrm{ext}_\Lambda(C)$. The sets $(A,X,B)$ clearly partition $V(\Lambda(G))$ and $X \cap V(G) = S$. Both $A$ and $B$ contain vertices of $G$ because $C$ is a fence. As $C$ corresponds to a closed Jordan curve, and contains no dummy vertices of $G$, there is no edge of $E(G) \cup E(\Lambda(G))$ with one endpoint in $A$ and the other endpoint in $B$. Therefore, $(A,X,B)$ is a co-separating triple. Now, we show that the nucleus $N_G[X \cap V(G)] = N_G[S]$ has diameter at most $|S|+4$ in $\Lambda(G)$. For any pair of vertices $u,v \in S$, there is a path along $C$ of length at most $|S|$, since $C$ has length $2|S|$. Any vertex in $N_G[S]$ has distance at most 1, in graph $G$, to a vertex of $S$. This translates to distance at most two in $G^\times$ to a vertex of $S$ (it is equal to two for a crossed edge). Hence, the nuclear diameter is at most $|S| + 4$.
\end{proof}


\section{1-Plane Graphs Without \texorpdfstring{\X}{X}-Crossings}\label{section: vertex connectivity of general 1-plane graphs}
\label{SEC:NOTFULL}

\begin{figure}
  \centering
  \begin{subfigure}[b]{0.45\textwidth}
    \centering
    \includegraphics{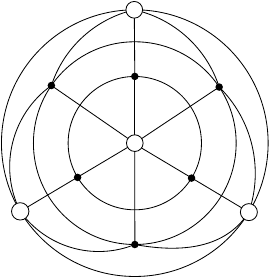}
    \caption{}
    \label{subfig: arrow embedding}
  \end{subfigure}
  \begin{subfigure}[b]{0.45\textwidth}
    \centering
    \includegraphics{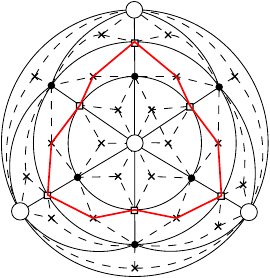}
    \caption{}
    \label{subfig: planarised arrow embedding}
  \end{subfigure}
  \caption{On the left is a 4-connected 1-plane graph $G$ where all crossings are arrow. A minimum vertex cut is marked with four white circles. On the right is a fence $C$ (bold edges) such that $S \subseteq N_{G^\times}(V_\times(C))$ and $V_\times(C) \subseteq N_{G^\times}[S]$.}
  \label{fig: sep set in an arrow embedding}
\label{fig:arrow_example}
\end{figure}

Unlike for full 1-plane graphs, we cannot extend \Cref{theorem: vertex connectivity planar graphs} to 1-plane graphs without \X-crossings. 
For example, consider Figure \ref{subfig: arrow embedding} showing a 1-plane graph where all crossings are arrow crossings. The graph is 4-connected and a minimum vertex cut $S$ is shown by vertices marked with white circles. One can verify that there is no cycle in $R(G)$ such that $V_\times(C) = S$. However, one may observe that there is a fence $C$ (shown with bold (red) edges in Figure \ref{subfig: planarised arrow embedding}) that skirts along $S$---as defined in \Cref{sec: building blocks}, this means that all $S$-vertices of are in the neighbourhood of $V_\times(C)$, and all vertices of $V_\times(C)$ are in the neighbourhood of $S$. As we now show in Theorem \ref{theorem: main theorem}, this is true for all 1-plane graphs without $\times$-crossings. (We will also use the graph in Figure \ref{subfig: arrow embedding} as a running example for the proof of Theorem \ref{theorem: main theorem}.)

\begin{theorem}\label{theorem: main theorem}
\label{thm:main}
Let $G$ be a 1-plane graph without \X-crossings and $S$ be a minimal vertex cut of $G$. Then there exists a fence $C \subseteq R(G)$ that skirts along $S$ such that $V(C) \cap V(G) \subseteq S$.
\end{theorem}

Before proceeding to give the full proof of \Cref{theorem: main theorem}, we sketch an outline here (\Cref{fig: schema} gives a schematic representation). In the first step, we fix two arbitrary flaps of $G \setminus S$, say $\phi_1$ and $\phi_2$, and add as many kite edges at crossings as possible, without merging $\phi_1$ with $\phi_2$. This step, which we call \textit{augmentation}, produces a graph $G_\text{aug}$ whose crossings are either full, almost-full, or arrow. In the next step, we add dummy vertices 
and edges to such crossings to
transform $G_\text{aug}$ into a full 1-pane graph $G_\boxtimes$. We then appeal to \Cref{theorem: vertex connectivity full 1-plane graph} to construct a fence $C_\boxtimes$ in $G_\boxtimes$, and show that $C_\text{aug} := C_\boxtimes$ is also a fence of $G_\text{aug}$. In the last step, we undo the augmentation step.
This will transform $G_\text{aug}$ back to $G$, and $C_\text{aug}$ to a fence $C$ of $G$ with the desired properties.

\begin{figure}%
  \begin{subfigure}[b]{0.32\textwidth}
    \includegraphics[width=\linewidth,page=2]{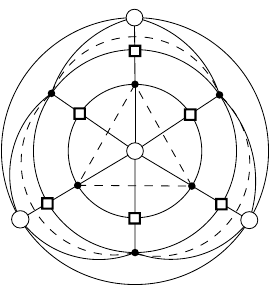}
    \caption{}
    \label{subfig: gbox}
  \end{subfigure}%
\hspace*{\fill}%
  \begin{subfigure}[b]{0.32\textwidth}
    \includegraphics[width=\linewidth,page=3]{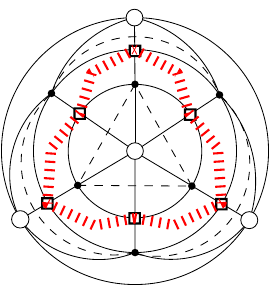}
    \caption{}
    \label{subfig: fence}
  \end{subfigure}%
\hspace*{\fill}%
  \begin{subfigure}[b]{0.32\textwidth}
    \includegraphics[width=\linewidth,page=4]{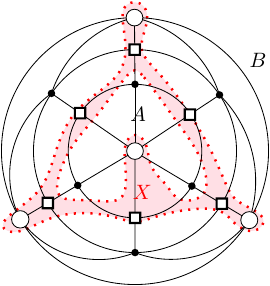}
    \caption{}
     \label{subfig: s and fence}
  \end{subfigure}%
  \caption{Finding a co-separating triple for the graph $G = G_\text{aug}$ from Figure~\ref{fig:arrow_example}.
(a) Graph $G_\boxtimes$; vertices in $D$ are marked with squares and base edges are dashed;
(b) Fence $C_\boxtimes \subseteq R(G_\boxtimes)$; here we have $C = C_\text{aug} = C_\boxtimes$.
(c) The resulting co-separating triple $(A,X,B)$ (face vertices are not shown).
}
\label{fig:arrow_example_contd}
\end{figure}

\begin{proof}[Proof of \Cref{theorem: main theorem}]
Fix two arbitrary flaps $\phi_1,\phi_2$ of $G\setminus S$. Now, we augment $G$ to $G_\text{aug}$ by adding as many kite-edges as possible without merging $\phi_1$ with $\phi_2$.

\begin{definition}[Augmentation]
Let $S$ be a minimal vertex cut and $\phi_1,\phi_2$ be two flaps of $G\setminus S$. The \emph{augmentation} of $G$ with respect to $\{S,\phi_1,\phi_2\}$ is the graph $G_\text{aug} := G_\text{aug}(S, \phi_1, \phi_2)$ obtained by the following iterative process: For any two consecutive endpoints $u,x$ of a crossing, if there is no kite edge $(u,x)$, and it could be added without connecting flaps $\phi_1,\phi_2$, then add the kite edge, update the flaps $\phi_1$ and $\phi_2$ (because they may have grown by merging with other flaps), and repeat.
\end{definition}

\begin{figure}
    \centering
    \includegraphics[width=0.5\linewidth]{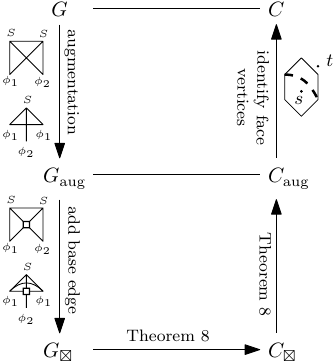}
    \caption{A schema illustrating the steps in the proof of \Cref{theorem: main theorem}, showing the transformations $G \rightarrow G_\text{aug} \rightarrow G_\boxtimes$ and $C_\boxtimes \rightarrow C_\text{aug} \rightarrow C$.}
    \label{fig: schema}
\end{figure}

In Lemma \ref{obs: augmentation of G and the structure of crossings}, we show that the crossings of $G_\text{aug}$ have some special properties (as illustrated in \Cref{fig: schema}).

\begin{lemma}\label{obs: augmentation of G and the structure of crossings}
If a crossing of $G_\text{aug} := G_\text{aug}(S, \phi_1, \phi_2)$ is not full, then one of the following is true:
\begin{enumerate}
    \item The crossing is almost-full with the spine vertices in $S$, one wing tip in $\phi_1$ and the other wing tip in $\phi_2$, or 
    \item The crossing is an arrow crossing with the tip in $S$, the tail in $\phi_2$, and the base vertices in $\phi_1$ (up to renaming $\phi_1$ and $\phi_2$).  
\end{enumerate}
\end{lemma}

\begin{proof}
Let $\{(u,v), (w,x)\}$ be a crossing of $G_\text{aug}$ that is not full. Since $G$ has no \X-crossing, there exists a kite edge; up to renaming vertices, we may assume this to be $(u,w)$. As a first case, suppose that 
$\{(u,x), (w,v)\} \subseteq E(G_\text{aug})$. As the crossing is not full, $(v,x) \notin E(G_\text{aug})$. Since the process of augmentation did not add the edge $(v,x)$, we must have $x \in \phi_1$ and $v \in  \phi_2$ (up to renaming flaps), and $\{u,w\} \subseteq S$ as they are adjacent to vertices in both flaps; such a crossing satisfies (1). Now for the second case, suppose that one of $\{(u,x), (w,v)\}$, say $(u,x)$, is not in $E(G_\text{aug})$. Up to renaming flaps, we must have $u\in \phi_1$ and $x\in \phi_2$. This implies $w\in S$ (since it is adjacent to both $u$ and $x$), and therefore $(w,v) \in E(G_\text{aug})$. Due to edge $(u,v)$ we have $v\in S\cup \phi_1$.  If $v\in S$, then $(x,v) \in E(G_\text{aug})$; the crossing is then an almost-full crossing that satisfies (1). If $v\in \phi_1$ then we have an arrow crossing that satisfies (2).
\end{proof}

We now convert $G_\text{aug}$ to a full 1-plane graph $G_{\boxtimes}$ as follows (see Figure~\ref{fig: Gref+}). At every almost-full crossing and every arrow crossing, replace the crossing point with a dummy vertex. Furthermore, at every arrow crossing, add a \emph{base edge} connecting the base vertices so that it forms a full crossing. Since every crossing of $G_\text{aug}$ is full, almost-full or arrow (\Cref{obs: augmentation of G and the structure of crossings}), all crossings of $G_\boxtimes$ are full. Let $D:=V(G_\boxtimes)\setminus V(G)$ be the set of newly added dummy vertices. Note that $S^+:=S\cup D$ is a vertex cut of $G_\boxtimes$ since no edge of $G_\boxtimes$ connects $\phi_1$ with $\phi_2$. (See Figure \ref{subfig: gbox}.)
 
As $G_\boxtimes$ is full 1-plane, we may apply Theorem \ref{theorem: vertex connectivity full 1-plane graph} to $G_\boxtimes$ and a subset of $S^+$ to obtain a fence $C_\boxtimes \subseteq R(G_\boxtimes)$ with $V(C_\boxtimes)\cap V(G_\boxtimes)\subseteq S\cup D$. As $C_\boxtimes$ does not visit dummy vertices, it is also a cycle in $R(G_\text{aug})$. (The only difference between $R(G_\text{aug})$ and 
$R(G_\boxtimes)$ is at each base edge: Here $R(G_\boxtimes)$
has an extra vertex at the crossing point created by the
base edge, and the four incident face vertices.) Hence, we let $C_\text{aug} := C_\boxtimes$. (See Figure \ref{subfig: fence}.)

\begin{lemma}\label{lemma: s,t in Caug}
$C_\text{aug}$ is a fence of $G_\text{aug}$, and there exist vertices $s \in \mathrm{int}_G(C_\text{aug})$ and $t \in \mathrm{ext}_G(C_\text{aug})$ where $s,t \in G \setminus S$.
\end{lemma}

\begin{proof}
As $C_\boxtimes = C_\text{aug}$ is a fence of $G_\boxtimes$, there exist $s' \in \mathrm{int}_{G_\boxtimes}(C_\text{aug})$ and $t' \in \mathrm{ext}_{G_\boxtimes}(C_\text{aug})$. We will show that if $s' \notin G \setminus S$, which is to say that $s' \in S \cup D$, then we can choose another vertex $s \in \phi_1 \cup \phi_2$ such that $s \in \mathrm{int}_G(C_\text{aug})$. (A symmetric argument works for $t'$ and $t$.) First, suppose that $s' \in D$. Then, by \Cref{obs: augmentation of G and the structure of crossings}, there exists an edge $(s',s) \in E(G^\times)$ where $s \in \phi_1 \cup \phi_2$ (also see Figure \ref{fig: Gref+}). Hence, we may assume that $s' \in S$. As each vertex of $S$ has a neighbour in $\phi_1 \cup \phi_2$ in $G_\boxtimes$, there exists a vertex $s'' \in \phi_1$ that is adjacent to $s'$. If $s'$ and $s''$ are on the same side of $C_\text{aug}$, then we simply choose $s := s''$, and we are done. Otherwise, $s'$ and $s''$ must be on opposite sides of $C_\text{aug}$, such that $(s',s'')$ is crossed at a vertex $d \in D$. (This is because $V_\times(C_\text{aug}) \subseteq S \cup D$.) Once again, by \Cref{obs: augmentation of G and the structure of crossings}, $s'$ has a neighbour $s$ in $G^\times$ such that $s \in \phi_1\cup \phi_2$ (also see Figure \ref{fig: Gref+}). As $(s',s) \in E(G^\times)$, $s$ must be on the same side of the cycle as $s'$. 
\end{proof}

We now transform the fence $C_\text{aug}$ of $\Lambda(G_\text{aug})$ into a fence $C$ of $\Lambda(G)$. First, remove the kite edges from $G_\text{aug}$ that were added to $G$ during augmentation. As this step unifies some faces of $G_\text{aug}$, some faces may contain more than one face vertex. In order to obtain $\Lambda(G)$, such face vertices must be identified so that each face has exactly one face vertex. This is done in two parts. Initially, we only identify face vertices on the fence, and then we identify the remaining face vertices. We describe this in detail now.  Initially, set $C := C_\text{aug}$. There are vertices $s,t \in G \setminus S$, one belonging to $\mathrm{int}_G(C)$ and the other to $\mathrm{ext}_G(C)$ (for the first iteration, this holds because of Lemma \ref{lemma: s,t in Caug}). Let $f_1$ and $f_2$ be two face vertices on $C$ that belong to the same face of $G$. Then, upon identifying $f_1$ and $f_2$, one can find a sub-cycle $C'$ (i.e., $E(C') \subseteq E(C)$) such that $s$ and $t$ are on opposite sides. Update $C$ to be the cycle $C'$, and repeat the above steps until no further face-vertices of $C$ can be identified. The cycle $C$ obtained after this process must be a fence, since it has $s$ and $t$ on opposite sides. Finally, we identify the remaining face-vertices---these are face vertices in $\mathrm{int}_\Lambda(C) \cup \mathrm{ext}_\Lambda(C)$. This gives us a fence $C$ of $\Lambda(G)$.

Now, we show that $C$ satisfies the requirements of \Cref{theorem: main theorem}. Since $V_\times(C) \subseteq S \cup D$, it follows that $V(C) \cap V(G) \subseteq S$. To show that $C$ skirts along $S$, we will need to prove that $V_\times(C) \subseteq N_{G^\times}[S]$ and $S \subseteq N_{G^\times}[(V_\times(C)]$. The former part is easy to show since $V_\times(C) \subseteq S \cup D$, and every vertex of $D$ is adjacent to a vertex of $S$ (\Cref{obs: augmentation of G and the structure of crossings,fig: Gref+}). This proves one part of \Cref{theorem: main theorem}. To prove the latter part, we show that $N_{G^\times}[C] \cap S$ is a vertex cut of $G$. As $S$ is a minimal vertex cut, this will imply that $S = N_{G^\times}[C] \cap S$  (\Cref{subfig: s and fence}). With this, the proof of \Cref{theorem: main theorem} will be complete.

\begin{figure}
  \centering
  \begin{subfigure}[b]{0.43\linewidth}
    \centering
    \includegraphics[width=0.4\textwidth,trim=0 0 120 0,clip]{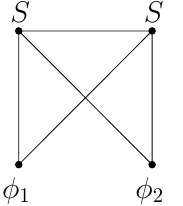}
    \raisebox{8ex}{$\mathbf{\rightarrow}$}
    \includegraphics[width=0.4\textwidth,trim=120 0 0 0,clip]{Figures/Planarising_almost_full_crossings.pdf}
  \end{subfigure}
\hspace*{\fill}
  \begin{subfigure}[b]{0.45\linewidth}
    \centering
    \includegraphics[width=0.45\textwidth,page=2,trim=0 0 160 0,clip]{Figures/Gref+.pdf}
    \raisebox{6ex}{$\mathbf{\rightarrow}$}
    \includegraphics[width=0.45\textwidth,page=2,trim=160 0 0 0,clip]{Figures/Gref+.pdf}
  \end{subfigure}
  \caption{ From $G_\text{aug}$ to
$G_\boxtimes$.  
Vertices in $D$ are grey squares, the base edge is dashed.
}
  \label{fig: Gref+}
\end{figure}

\begin{lemma}\label{claim: S' vertex cut}
The set $N_{G^\times}[C] \cap S$ separates $\mathrm{int}_G(C)$ and $\mathrm{ext}_G(C)$ in $G$.  
\end{lemma}

\begin{proof}
Let $P$ be any arbitrary $(a,b)$-path in $G$, for any pair of vertices $a \in \mathrm{int}_G(C)$ and $b \in \mathrm{ext}_G(C)$. To prove \Cref{claim: S' vertex cut}, it is sufficient to show that $P$ contains a vertex of $N_{G^\times}[C] \cap S$. Let $(r,w,s)$ be three consecutive vertices on $P^\times$ (the path obtained from $P$ by adding a dummy vertex wherever an edge of $P$ is crossed) such that $r \in \mathrm{int}_\Lambda(C)$, $w \in V(C)$, and $s \in V(C) \cup \mathrm{ext}_\Lambda(C)$; these vertices exist since $C$ is a closed Jordan curve separating $a$ and $b$. We will show that at least one of $\{r,w,s\}$ is a vertex of $S$. If $w \in S$, then we are already done. Otherwise, $w$ must be a dummy vertex (because $V_\times(C) \subseteq S \cup D$) on an edge $(r,s) \in E(G)$. The crossing at $(r,s)$ cannot be a full crossing since $C$ does not contain dummy vertices of full crossings ($C$ is derived from $C_\boxtimes$). By Lemma \ref{obs: augmentation of G and the structure of crossings}, the only remaining possibilities are that $(r,s)$ is either an edge of an almost full crossing or an arrow crossing in $G_\text{aug}$. If $(r,s)$ is part of an almost full crossing of $G_\text{aug}$, or part of an arrow crossing with $r$ and $s$ as the tip and tail, then we are done since at least one of $\{r,s\}$ is a vertex of $S$ (\Cref{obs: augmentation of G and the structure of crossings}). The remaining case is when $r$ and $s$ are base vertices. We will show that this situation cannot occur. Consider the full crossing in $G_\boxtimes$ that we get by adding the base edge. The cycle $C_\boxtimes \subseteq R(G_\boxtimes)$ does not visit any of the four face vertices of $R(G_\boxtimes)$, because otherwise it would have to continue via the dummy-vertex (but $C_\boxtimes$ does not visit dummy-vertices of full crossings) or via one of the base-vertices in $\Phi_1$ (but $G_\boxtimes$-vertices of $C_\boxtimes$ belong to $S\cup D$). 
Hence, there is no part of $C_\boxtimes = C_\text{aug}$ drawn inside the full crossing. Therefore, the simple Jordan arc that one can draw from $r$ to $s$ in $G_\text{aug}$ without intersecting $C_\text{aug}$ can also be drawn in $G$ without intersecting $C$. This goes on to show that $\{r,s\} \subseteq \mathrm{int}_\Lambda(C)$---a contradiction.  
\end{proof}
This completes the proof of \Cref{theorem: main theorem}.
\end{proof}

Similar to \Cref{thm: co-separating triple plane graph,thm: co-separating triple full 1-plane}, we now show that the existence of a fence $C$ as in \Cref{theorem: main theorem} implies a co-separating triple $(A,X,B)$ of $(\Lambda(G),G))$ with small nuclear diameter.

\begin{theorem}\label{thm: co-separating triple and nuclear diameter}
Let $G$ be a 1-plane graph without $\times$-crossings. For any minimal vertex cut $S$ of $G$, there exists a co-separating triple $(A,X,B)$ of $(\Lambda(G),G))$ such that $X \cap V(G) = S$ and the nuclear diameter is at most $4|S|$.
\end{theorem}

\begin{proof}
From Theorem \ref{theorem: main theorem}, we know that there exists a fence $C \subseteq R(G)$ such that $S \subseteq N_{G^\times}[V_\times(C)]$, $V_\times(C) \subseteq N_{G^\times}[S]$ and $V_\times(C)\cap V(G) \subseteq S$. We use the cycle $C$ to partition the vertices of $\Lambda(G)$ into three sets $A$, $X$ and $B$ as follows. Let $X = S \cup V(C)$, and $A$ and $B$ be the remaining vertices of $\Lambda(G)$ in $\mathrm{int}_\Lambda(C)$ and $\mathrm{ext}_\Lambda(C)$ respectively. Clearly, $(A,X,B)$ is a partition of $V(\Lambda(G))$ and $X \cap V(G) = S$. We will show that $(A,X,B)$ is a co-separating triple of $\Lambda(G)$. As $C$ is a fence, Condition \ref{c1} (each of $A$, $X$, $B$ contain a vertex of $G$) and Condition \ref{c2} (no edge in $\Lambda(G)$ between $A$ and $B$) follow immediately. We are left to show Condition \ref{c3}. This requires that for every edge $e \in G \setminus (X \cap V(G))$, all vertices of $e^\times \cap \Lambda(G)$ belong entirely to $A \cup X$, or to $B \cup X$. This condition is violated only when there exists an edge $e = (a,b)$ with $a \in A$ and $b \in B$. But, no such edge exists by Lemma \ref{claim: S' vertex cut}. This proves that $(A,X,B)$ is a co-separating triple of $(\Lambda(G),G)$.

We now prove that the nucleus of $(A,X,B)$ has diameter at most $4|S|$. For any pair of vertices $u,v$ in the nucleus $N_G[X \cap V(G)] = N_G[S]$, let $\{s_u, s_v\} \subseteq S$ be vertices that are closest in $\Lambda(G)$ to $u$ and $v$ respectively. Consider a walk from $s_u$ to $s_v$ along cycle $C$ as $\{s_u\} \cup \{x_0,f_1,x_2,f_3,\dots,x_{2t}\} \cup \{s_v\}$, where all vertices $x_{2i}$ belong to $V_\times(C)$, and all vertices $f_{2i+1}$ are face vertices on $C$. (If $s_u \in V(C)$, then $x_0 = s_u$; else there exists an edge $(s_u,x_0)$ in $\Lambda(G)$ by $S\subseteq N_{G^\times}[V_\times(C)]$. Likewise, $x_{2t} = s_v$ or $(x_{2t},s_v) \in E(\Lambda(G))$.) We use this to define another walk that detours at each vertex $x_{2i}$ to reach the nearest $S$-vertex. To this end, we associate a vertex $s_{2i}$ for each $x_{2i}$, for $i \in \{1,\dots,t-1\}$, as follows: if $x_{2i} \in S$, set $s_{2i} := x_{2i}$; else set $s_{2i}$ to be the $S$-vertex neighbouring $x_{2i}$ in $\Lambda(G)$ (this exists by $V_\times(C)\subseteq N_{G^\times}[S]$). By detouring at each vertex $x_{2i}$ to reach $s_{2i}$, for $i \in \{1,\dots,t-1\}$, we get the following walk: $\{s_u\} \cup \{x_0,f_1,(x_2,s_2,x_2),f_3,\dots,f_{2i-1},(x_{2i},s_{2i},x_{2i}),f_{2i+1},\dots,f_{2t-1},x_{2t} \} \cup \{s_v\}$. We now prune this walk by removing all vertices that occur between the same $S$-vertex. After pruning, we get a walk where each $S$-vertex occurs at most once, and there are at most 3 vertices between any two $S$-vertices. Therefore, the length of the walk from $s_u$ to $s_v$ is at most $4(|S|-1)$. As $u \in N_G[S]$, either $u = s_u$, or $(u,s_u) \in E(G)$. Therefore, the distance in $\Lambda(G)$ between $u$ and $s_u$ is at most two (it is two when $(u,s_u)$ is a crossed edge), and similarly for $v$ and $s_v$. Hence, there is a walk from $u$ to $v$ of length at most $4(|S|-1) + 4 = 4|S|$.  
\end{proof}


\section{Computing Vertex Connectivity in Linear Time}\label{section: computing vertex connectivity in linear time}
\label{SEC:LINEAR}

By \Cref{thm: co-separating triple and nuclear diameter}, if a 1-plane graph $G$ without $\times$-crossings has a vertex cut of size at most $k$, then there exists a co-separating triple $(A,X,B)$ of $(\Lambda(G),G)$ such that $|X \cap V(G)| \leq k$. Conversely, by \Cref{obs: co-sep triple}, if $(A,X,B)$ is any co-separating triple of $(\Lambda(G),G)$ such that $|X \cap V(G)| \leq k$, then $G$ has a vertex cut of size at most $k$. Therefore, the problem of computing a minimum vertex cut of $G$ is equivalent to the problem of finding a co-separating triple that minimises $|X \cap V(G)|$. The objective of this section is to use this correspondence to prove the following theorem:

\begin{theorem}\label{thm: linear time}
    The vertex connectivity of a 1-plane graph without $\times$-crossings can be computed in linear time.
\end{theorem}

To facilitate searching for a co-separating triple $(A,X,B)$ of $(\Lambda(G),G)$ with minimum $|X \cap V(G)|$, we make use of the nuclear diameter property of \Cref{thm: co-separating triple and nuclear diameter}: corresponding to every minimum vertex cut of $G$ (which has size at most 7 \cite{Bodendiek1983BemerkungenZE,pach1997graphs}), there exists a co-separating triple with nuclear diameter at most $w:= 4 \times 7 = 28$. Then, we construct various planar graphs $\Lambda_i$ of radius $O(w)$, where $i \in \{0,\dots,d\}$ for some finite integer $d$, whose vertices are subsets of the vertices of $G$ and together cover $\Lambda(G)$. These graphs $\Lambda_i$ are such that every co-separating triple $(A,X,B)$ of $(\Lambda(G),G))$ can be projected onto a co-separating triple $(A_i,X,B_i)$ of $(\Lambda_i,G)$, for some $i \in \{0,\dots,d\}$ (notice that $X$ remains unchanged). Conversely, every co-separating triple $(A_i,X,B_i)$ of $(\Lambda_i,G)$ can be lifted into a co-separating triple $(A,X,B)$ of $(\Lambda(G),G))$. By this, the problem of finding a co-separating triple $(A,X,B)$ of $(\Lambda(G),G))$ with minimum $|X \cap V(G)|$ becomes equivalent to the problem of finding a co-separating triple $(A_i,X,B_i)$ of $(\Lambda_i,G))$ with minimum $|X \cap V(G)|$, among all values of $i \in \{0, \dots, d\}$. As the graphs $\Lambda_i$ are planar with radius $O(w) \subseteq O(1)$, they have bounded treewidth. This leaves at our disposal the well-known framework of Monadic Second Order Logic (MSOL) and Courcelle's theorem, using which we can find such a co-separating triple in overall linear time.

\subsection{Graphs \texorpdfstring{$\Lambda_i$}{Lambda(i)}}
As a first step, we perform a breadth-first search (BFS) in $\Lambda(G)$ starting at an arbitrary root vertex; let $T$ be the resulting BFS-tree. For $j \in \{0,\dots,d\}$, let $V_j$ (the \emph{$j$th layer}) be the vertices at distance $j$ from the root, and let $d$ be the largest index where $V_j$ is non-empty. Define $V_j:=\emptyset$ for any index $j<0$ or $j>d$. For any $a\leq b$, let $\Lambda[V_a\cup \dots \cup V_b]$ denote the subgraph of $\Lambda(G)$ induced by $V_a\cup \dots \cup V_b$. For any index $j$, let $E_j$ be the set of edges with both endpoints in $V_j$, and let $E_{j,j+1}$ be the set of edges that connect $V_{j}$ and $V_{j+1}$; these sets cover all edges of $\Lambda(G)$. Let $w := 4 \times 7 = 28$ be the maximum value of the nuclear diameter of any co-separating triple that corresponds, as in \Cref{thm: co-separating triple and nuclear diameter}, to a minimum vertex cut of $G$. Now, we use the BFS layers $V_0, \dots, V_d$ to construct graphs $\Lambda_i$ as follows.

\begin{figure}
    \centering
    \includegraphics[scale = 0.6]{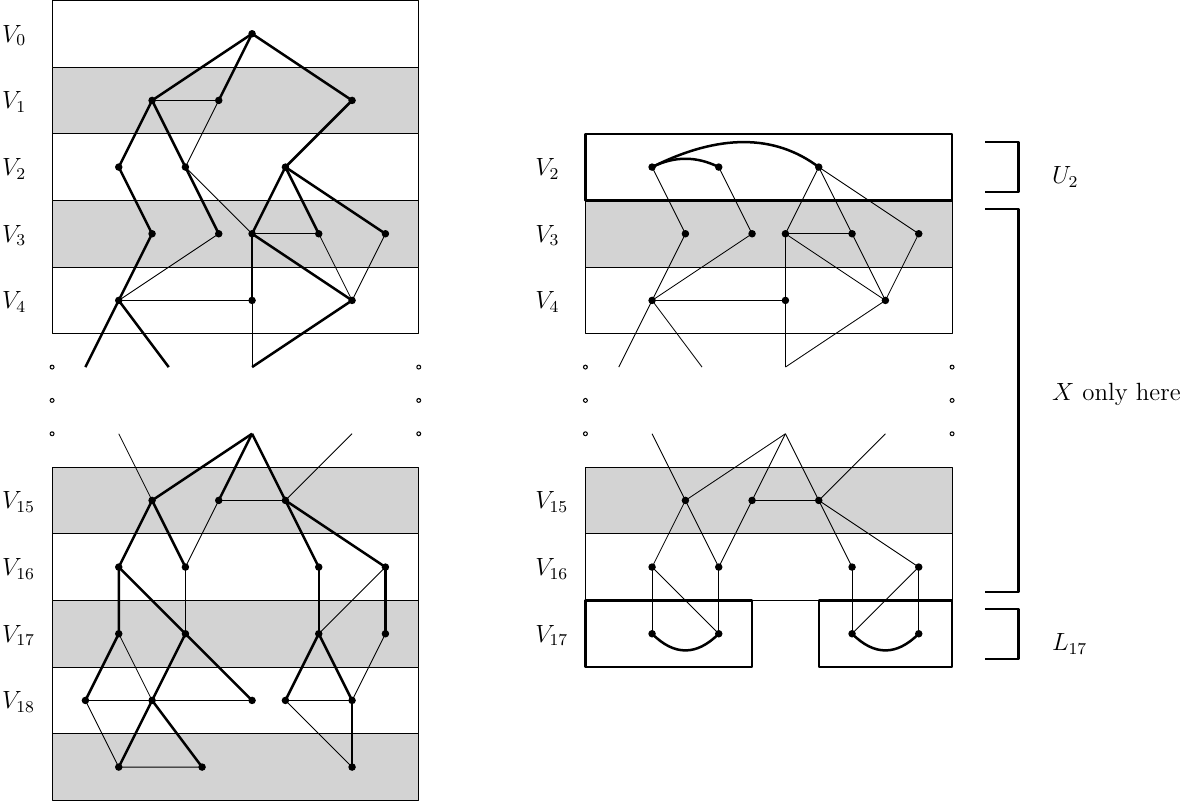}
    \caption{ Construction of $\Lambda_3$ for $w=14$.  Edges in $U_2$ and $L_{17}$ are bold.}
    \label{fig: BFS tree and BFS Tree G_i (alternate)}
\end{figure} 

\begin{lemma}
\label{claim:U and L}
For $i\in \{0,\dots,d\}$, 
there exists a graph $\Lambda_i := \Lambda[V_{i-1}{\cup} \dots {\cup} V_{i+w+1}] \cup (V_{i-1},U_{i-1}) \cup (V_{i+w+1},L_{i+w+1})$, where $(V_{i-1},U_{i-1})$ and $(V_{i+w+1},L_{i+w+1})$ are graphs such that:
\begin{enumerate}
\item For any $u,v \in V_{i-1}$, there is a $(u,v)$-path in $(V_{i-1}, U_{i-1})$ if and only if there is a $(u,v)$-path in $\Lambda[V_{0}\cup \dots \cup V_{i-1}]$.   
\item For any $u,v \in V_{i+w+1}$, there is a $(u,v)$-path in $(V_{i+w+1}, L_{i+w+1})$ if and only if there is a $(u,v)$-path in $\Lambda[V_{i+w+1}\cup \dots \cup V_{d}]$.    
\item $\Lambda_i$ is planar and has radius at most $w{+}2$.
\end{enumerate}
 The total time to compute edge sets $\{U_i,L_i\}$ for all $i \in \{0,\dots,d\}$ is $O(|V(\Lambda(G))|)$.
\end{lemma}

\begin{proof}
It is easy to show the existence of $\Lambda_i$ (as argued in \cite{eppstein}): Contract every edge of $\Lambda[V_0 \cup \dots \cup V_{i-1}] \cup \Lambda[V_{i+w+1} \cup \dots \cup V_{d}]$ that has at most one endpoint in $V_{i-1} \cup V_{i+w+1}$. Then let $U_{i-1}$ and $L_{i+w+1}$ be the sets of edges within $V_{i-1}$ and $V_{i+w+1}$ that result from this process. Although this description is simple, it is not obvious how one could implement contraction in overall linear time; hence we give an alternate approach. 

It is relatively easy to construct the graph $(V_{i-1}, U_{i-1})$. Pick an arbitrary vertex $v$ in $V_{i-1}$, and let $U_{i-1} := \{(v,w): w \in U_{i-1}\}$ (\Cref{fig: BFS tree and BFS Tree G_i (alternate)}). In consequence, all vertices of $V_{i-1}$ can be connected within $U_{i-1}$; this is appropriate since they can all be connected using only layers $V_0 \cup \dots \cup V_{i-1}$. We refer to $v$ as the \textit{root of $\Lambda_i$}. 

We now show how to construct graphs $(V_j,L_j)$. We start in reverse order, from $j=d$ down to $j=0$. For any vertex $w \in \Lambda(G)$, let $p(w)$ be the parent of $w$ in the BFS-tree. For the last layer, simply set $L_d := E_d$. For $j<d$, we first build a graph $(V_j, \hat{L}_j)$, where $\hat{L}_j := E_j \cup \{(v,p(w)): v \in V_j \text{, } w \in V_{j+1} \text{ and } (v,w) \in E_{j,j+1}\} \cup \{(p(v), p(w)): (v,w) \in L_{j+1}\}$. Then we simplify $(V_j,\hat{L}_j)$ by computing the underlying simple graph $(V_j,L_j)$.

\begin{claim}
Graphs $(V_j,U_j)$ and $(V_j,L_j)$ can be obtained from $\Lambda[V_0\cup \dots \cup V_j]$ and $\Lambda[V_j\cup \dots \cup V_d]$ via a sequence of edge-contractions and simplifications.
\end{claim}

\begin{proof}  
It is trivial to show this for $(V_j, U_j)$: contract every edge of $\Lambda[V_0 \cup \dots \cup V_{j-1}]$ yielding one supernode,
and then contract this supernode into the root of $\Lambda_j$. To show that this is true for $(V_j, L_j)$ as well, we do induction on $j$. As $L_d = E_d$, this clearly holds for $j=d$, so assume that $j<d$. We can obtain graph $(V_j,\hat{L}_j)$ from $\Lambda[V_j\cup \dots \cup V_d]$ as follows. First perform all the edge-contractions and simplifications necessary to convert
subgraph $\Lambda[V_{j+1}\cup \dots \cup V_d]$ into $(V_{j+1},L_{j+1})$;
this is possible by induction. Now, for every vertex $w\in V_{j+1}$,
contract $w$ into its parent $p(w)$. For any neighbour $v$ of $w$, if $v\in V_j$,
then $(v,p(w)) \in \hat{L}_j$, and if $v\in V_{j+1}$, then $(p(v),
p(w)) \in \hat{L}_j$.  Vice versa, every edge of $\hat{L}_j$ is either an edge of $\Lambda[V_j]$ or can obtained by contracting a vertex $w \in V_{j+1}$ into its parent $p(w)$. Upon simplifying 
$(V_j,\hat{L}_j)$, we get $(V_j,L_j)$.
\end{proof}

With this, Lemma~\ref{claim:U and L}(1) and \ref{claim:U and L}(2) are satisfied since 
connected components are not changed when doing edge-contractions and simplifications. 
Since both $U_{i-1}$ and $L_{i+w+1}$ can be seen as being obtained via contractions, $\Lambda_i$ is planar. To show Lemma~\ref{claim:U and L}(3), we must establish the radius bound. Any vertex $v\in \Lambda_i$ has distance at most $w{+}2$ to the root of $\Lambda_i$, because we can go upward from $v$ using edges of the BFS-tree at most $w{+}1$ times until we reach a vertex $v$ in $V_{i-1}$, from where we can reach the root of $\Lambda_i$ using at most a single edge.

It remains to analyze the run-time. Constructing $U_{i}$ takes $O(|V_i|)$ time, so the total time to construct all $U_i$ is $O\left(\sum_{i=0}^d |V_i|\right) = O(|V(\Lambda(G))|)$. Constructing $\hat{L}_j$ is proportional to the number of edges that it contains,
which, by its construction, is at most $|E_j| + |E_{j,j+1}| + |L_{j+1}|$. Graph $(V_{j+1},L_{j+1})$ is planar and simple, so $|L_{j+1}|\in O(|V_{j+1}|)$.  The time it takes to simplify the graph $(V_j, \hat{L}_j)$ is $O(|V_j| + |\hat{L_j}|)$. Therefore, the time to construct $L_j$ is $O(|E_j|+|E_{j,j+1}| + |V_j|+ |V_{j+1}|)$. Summing over all $j$, the run-time is
$O\left (\sum_{j=0}^{d} \left(|E_j|+|E_{j,j+1}| + |V_j|+ |V_{j+1}|\right)\right)
\subseteq O\left(|E(\Lambda(G))| + |V(\Lambda(G))| \right) \subseteq O(|V(\Lambda(G))|)$.
\end{proof}

\subsection{Co-separating Triples in \texorpdfstring{$\Lambda_i$}{Lambda(i)}} 

Now we show that every co-separating triple $(A,X,B)$ of $(\Lambda(G),G)$ with nuclear diameter at most $w$ can be projected onto a co-separating triple $(A_i,X,B_i)$ of $(\Lambda_i,G)$, and conversely, every co-separating triple $(A_i,X,B_i)$ of $(\Lambda_i,G)$ can be lifted into a co-separating triple $(A,X,B)$ of $(\Lambda(G),G)$. For ease of reference, we repeat the definition of a co-separating triple here.

\CoSeparatingTriple*

\begin{lemma}
If there exists a co-separating triple $(A, X, B)$ of $(\Lambda(G), G)$ with nuclear
diameter at most $w$, then there exists a co-separating triple $(A_i,X,B_i)$ of $(\Lambda_i, G)$, for some $i \in \{0,\dots,d\}$, where $X \subseteq \Lambda[V_i \cup \dots \cup V_{i+w}]$.   
\end{lemma}

\begin{proof}
As the nuclear diameter of $(A,X,B)$ is at most $w$, all vertices of the nucleus belong to $V_{i} \cup \dots \cup V_{i+w}$ for some $i\in \{1,\dots,d\}$. Consider the graph $\Lambda_i$ for this choice of $i$. Let $A_i = A \cap V(\Lambda_i)$ and $B_i = B \cap V(\Lambda_i)$. We will show that $(A_i,X,B_i)$ is a co-separating triple of $(\Lambda_i, G)$. Since the nucleus contains some vertices of $A \cap V(G)$ and $B \cap V(G)$ (\Cref{obs: nucleus elements}), each of $A_i$, $X$ and $B_i$ contains a vertex of $G$. Hence, Condition \ref{c1} holds. Since $A_i \subseteq A$ and $B_i \subseteq B$, for any edge $e \in E(G \setminus (X \cap V(G)))$, any vertex of $e^\times \cap \Lambda_i$ that belonged to $A$, $X$ or $B$ now belongs to $A_i$, $X$ or $B_i$, respectively; hence Condition \ref{c3} holds as well. 
Condition \ref{c2} clearly holds
for all edges in $E(\Lambda_i) \cap E(\Lambda(G))$. Any edge in $E(\Lambda_i) \setminus E(\Lambda(G))$ must belong to $U_{i-1}$ or $L_{i+w+1}$. By \Cref{claim:U and L}, every component of $(V_{i-1}, U_{i-1}) \cup (V_{i+w+1}, L_{i+w+1})$ is a projection of some component of $\Lambda[V_0 \cup \dots \cup V_{i-1}] \cup \Lambda[V_{i+w+1} \cup \dots \cup V_{d}]$. By Condition \ref{c2} applied to $(A,X,B)$, all vertices in the same component of $\Lambda[V_0 \cup \dots \cup V_{i-1}] \cup \Lambda[V_{i+w+1} \cup \dots \cup V_{d}]$ either belong to $A$ or to $B$. Therefore, all vertices in the same component of $(V_{i-1}, U_{i-1}) \cup (V_{i+w+1}, L_{i+w+1})$ belong to $A_i$ or $B_i$. 
\end{proof}

\begin{lemma}
If $(A_i,X,B_i)$ is a co-separating triple of $(\Lambda_i, G)$ where $X \subseteq V_i \cup \dots \cup V_{i+w}$, then there exist sets $A$ and $B$ such that $(A, X, B)$ is a co-separating triple of $(\Lambda(G), G)$.
\end{lemma}

\begin{proof}
We use the existing co-separating triple $(A_i,X,B_i)$ to build a co-separating triple $(A,X,B)$ for $(\Lambda(G), G)$. Initially, all vertices in $V_{i} \cup \dots \cup V_{i+w}$ that belong to $A_i$ and $B_i$ are added to $A$ and $B$ respectively. All the other remaining vertices of $\Lambda(G)$ are in some component of $\Lambda[V_0 \cup \dots \cup V_{i-1}]$ or $\Lambda[V_{i+w+1} \cup \dots \cup V_{d}]$. 
By Condition \ref{c2} applied to $(A_i,X,B_i)$, and the assumption that $X \subseteq V_{i} \cup \dots \cup V_{i+w}$, all vertices in the projection of these components belong to $A_i$ or $B_i$. We can therefore unambiguously assign each vertex of $\Lambda[V_0 \cup \dots \cup V_{i-1}] \cup \Lambda[V_{i+w+1} \cup \dots \cup V_{d}]$ to $A$ or $B$ depending on whether vertices in their projection belong to $A_i$ or $B_i$ respectively.  

We must now argue that $(A,X,B)$ is a co-separating triple. Each of $A$, $X$ and $B$ contains a vertex of $G$ because $A_i \subseteq A$ and $B_i \subseteq B$; hence Condition \ref{c1} holds. Next, we show that Condition \ref{c2} holds. Any edge of $\Lambda(G)$ is either an edge of $\Lambda[V_{i} \cup \dots \cup V_{i+w}]$ or an edge in some component of $\Lambda[V_0 \cup \dots \cup V_{i-1}] \cup \Lambda[V_{i+w+1} \cup \dots \cup V_{d}]$. In the first case, the endpoints of an edge cannot belong to $A$ and $B$ since $(A_i,X,B_i)$ was itself a co-separating triple. For the second case, note that by construction, all vertices in the same component of $\Lambda[V_0 \cup \dots \cup V_{i-1}] \cup \Lambda[V_{i+w+1} \cup \dots \cup V_{d}]$ belong to $A$ or $B$. Hence, Condition \ref{c2} holds. We are left with showing that Condition \ref{c3} holds. Let $e \in E(G \setminus (X \cap V(G)))$. If no vertex of $e^\times$ belongs to $\Lambda_i$, then all of them belong to the same component of $\Lambda[V_0 \cup \dots \cup V_{i-1}] \cup \Lambda[V_{i+w+1} \cup \dots \cup V_{d}]$, and therefore belong $A$ or $B$. So, we can assume that $e^\times \cap \Lambda_i \neq \emptyset$. Assume that all vertices of $e^\times \cap \Lambda_i$ belong to, say, $A\cup X$ (by Condition \ref{c3} on $(A_i,X,B_i)$). For any vertex $z\in e^\times$ that is not in $\Lambda_i$, there exists a vertex $z' \in e^\times$ such that the subpath in $e^\times$ from $z$ to $z'$ belongs to some component of $\Lambda[V_0\cup \dots \cup V_{i-1}]$ or $\Lambda[V_{i+w+1}\cup \dots\cup V_{d}]$. Therefore $z$ belongs to the same set as $z'$, i.e., $A$. 
Hence, Condition \ref{c3} holds, establishing that $(A,X,B)$ is a co-separating triple.
\end{proof}

\subsection{Graphs \texorpdfstring{$\Lambda_i^+$}{Lambda(i)+}}

To simplify finding co-separating triples in $\Lambda_i$, we construct a graph $\Lambda_i^+$ so that we eliminate the requirement to test Condition \ref{c3} for co-separating triples. To construct $\Lambda_i^+$, begin with graph $\Lambda_i$, and for every edge $(v,w) \in E(G)$, if $(v,w)$ is crossed at a point $c$, and $\{v,c,w\} \subseteq V(\Lambda_i)$, add edge $(v,w)$ to $\Lambda_i^+$. (Note that $\Lambda_i^+$ need not be planar; this will not pose problems.)

\begin{lemma}
Let $(A_i,X,B_i)$ be a partition of $V(\Lambda_i)$ such that $X \subseteq V_i \cup \dots \cup V_{i+w}$. Then $(A_i, X, B_i)$ is a co-separating triple of $\Lambda_i$ if and only if it satisfies Conditions \ref{c1} and \ref{c2} of co-separating triples (\Cref{def: co-separating triple}) for $(\Lambda_i^+,G)$.
\end{lemma}

\begin{proof}
Suppose that $(A_i,X,B_i)$ is a co-separating triple of $(\Lambda_i,G)$. Then Condition \ref{c1} is trivially satisfied for $(\Lambda_i^+,G)$. Condition \ref{c2} holds for all edges in $E(\Lambda_i^+) \cap E(\Lambda_i)$. The remaining edges of $\Lambda_i^+$ are all edges of $G$. By Condition \ref{c3} on $(\Lambda_i,G)$, no edge of $G$ can have one endpoint in $A$ and the other in $B$. 

For the other direction, we need to show that if $(A_i, X, B_i)$ satisfies Conditions \ref{c1} and \ref{c2} for $(\Lambda_i^+,G)$, then $(A_i, X, B_i)$ is a co-separating triple of $(\Lambda_i,G)$. Condition \ref{c1} holds because $V(\Lambda_i) = V(\Lambda_i^+)$, and Condition \ref{c2} holds because $E(\Lambda_i) \subseteq E(\Lambda_i^+)$. The only non-trivial aspect here is to show that Condition \ref{c3} holds. Suppose, for contradiction, that there is an edge $(a,b)$ of $G \setminus (X \cap V(G))$ such that $a \in A_i$ and $b \in B_i$. Such a contradiction can arise only if $(a,b) \notin E(\Lambda_i^+)$. This happens if $(a,b)$ is crossed at a point $c$, and $\{a,b\} \subseteq V_{i-1} \cup V_{i+w+1}$ with $c$ belonging to the adjacent layer not in $\Lambda_i$. Let us assume that $c$ belongs to $V_{i+w+2}\cup \dots \cup V_d$; the other case is symmetric.
Then path $a$-$c$-$b$ connects $a$ and $b$ within layers $V_{i+w+1}\cup \dots \cup V_d$. By
\Cref{claim:U and L}, $a$ and $b$ are in the same component of $(V_{i+w+1},L_{i+w+1})$.
This would imply that $\{a,b\} \subseteq A_i$ or $\{a,b\} \subseteq B_i$, contradicting our assumption.
\end{proof}

To determine whether there exist sets $(A_i,X,B_i)$ in $\Lambda_i^+$ that satisfy Conditions \ref{c1} and \ref{c2}, we construct a tree decomposition of $\Lambda_i^+$ with small width (defined below). Thereafter, we show that one can use monadic second order logic (MSOL) to give a formulaic description for the existence of sets $(A_i,X,B_i)$ in $\Lambda_i^+$ as required. Then, using Courcelle's famous theorem, we can evaluate the MSOL formula in linear time.  

\subsubsection{Tree Decomposition of \texorpdfstring{$\Lambda_i^+$}{Lambda(i)+}}

First, we define tree decomposition and tree width. The reader may refer to \cite{BK08} for an overview of the algorithmic implications for graphs of bounded treewidth (in particular, many NP-hard problems become polynomial).

\begin{definition}[Tree Decomposition]
A \emph{tree decomposition} of a graph $G$ is a tree $\calT$ whose
nodes are associated with subsets of $V(G)$ (`\emph{bags}') such that
\begin{enumerate}
    \item Every vertex $v$ of $G$ is \emph{covered}, i.e., $v$ belongs to at least one bag of $\calT$;
    \item Every edge $(u,v)$ of $G$ is \emph{covered}, i.e., there exists a bag $Y\in V(\calT)$ with $u,v\in Y$;
    \item For every vertex $v$ of $G$, the set $\calT[v]$ of bags containing $v$ is \emph{contiguous}, i.e., forms a connected subtree of $\calT$.
\end{enumerate}
The \emph{width} of a tree decomposition is $\max_{Y\in V(\calT)} |Y|-1$,
and the \emph{treewidth} of a graph $G$ is the smallest width of a tree
decomposition of $G$. 
\end{definition}

Let $\calT_i$ be a tree decomposition of $\Lambda_i$ with $|\calT_i|$ bags. We can create a tree decomposition ${\cal T}_i^+$ of $\Lambda_i^+$ as follows. Begin with $\calT_i$. For any bag $Y$ and any dummy vertex $c\in Y$, add to $Y$ all neighbours of $c$ in $\Lambda_i$.

\begin{claim}\label{claim: construct calT+}
$\calT_i^+$ is a tree decomposition of $\Lambda_i^+$ such that $width(\calT_i^+) \leq 5 \cdot width(\calT_i) + 4$, 
and $\calT_i^+$ can be computed from $\calT_i$ in time $O(width(\calT_i) \cdot |\calT_i|)$.
\end{claim}

\begin{proof}
We first show that $\calT_i^+$ is a tree decomposition of $\Lambda_i^+$. Every vertex of $\Lambda_i^+$ is trivially covered, since we started with a tree decomposition of $\Lambda_i$. We now show that every edge $(u,v)$ of $\Lambda_i^+$ is covered. If $(u,v)\in \Lambda_i$, then it was already covered by $\calT_i$.  If $(u,v)\notin \Lambda_i$, then it is an edge of $G$ that is crossed at a dummy vertex $c$ such that $\{u,c,v\} \subseteq V(\Lambda_i)$. Hence, there is a bag of $\calT_i$ that contains $c$, and both $u,v$ are added to this bag. Therefore, every edge of $\Lambda_i^+$ is covered by some bag of $\calT_i^+$. We are left with showing that for any vertex $v\in \Lambda_i$, the set $\calT_i^+[v]$ of bags containing $v$ is contiguous. Notice that vertex $v$ is added to more bags only if it is a neighbour of some dummy vertex $c$. Since $(v,c) \in E(\Lambda_i)$, the intersection $\calT_i[v]\cap \calT_i[c]$ is non-empty. As $v$ is added to all bags containing $c$, the new set of bags containing $v$ is again contiguous.
  
The remaining parts of \Cref{claim: construct calT+} are easy to show. Since dummy vertices have 4 neighbours in $\Lambda(G)$, we add at most 4 new vertices for each vertex in a bag. As the number of vertices in each bag increases at most 5-fold, $(width(\calT_i)^+ + 1) \leq 5 \times (width(\calT_i)+1)$; hence $width(\calT_i^+) \leq 5\cdot width(\calT_i)+4$. 
Also, we spend constant time per vertex per bag, and so the time to compute $\calT_i^+$ is $O(width(\calT_i) \cdot |\calT_i|)$.
\end{proof}

\subsection{Monadic Second-Order Logic}\label{sec: msol}

Monadic second-order logic (MSOL) is a logic where one can quantify over elements, over sets of elements (but not over sets of pairs, triples, etc.), and one can test membership of elements in sets. Here, we restrict ourselves to the application of MSOL for formulating graph decision problems. (For a more detailed treatment, the interested reader may refer to \cite{CE12}.). In MSOL for graphs, we have a binary relation $E(x,y)$ that is true if and only if $(x,y)$ is an edge. We may also have a finite set of unary relations on vertices; in our application we will need $V_j(x)$ that is true if and only if vertex $x$ belongs to layer $V_j$, and $V_G(x)$ that is true if and only if vertex $x$ belongs to $V(G)$. We are also allowed to use quantifiers $\exists,\forall$ over vertex-sets, but not over edge-sets.

Recall that we would like to find sets $(A_i,X,B_i)$ that partition $V(\Lambda_i^+)$, satisfy Conditions \ref{c1} and \ref{c2} of co-separating triples (\Cref{def: co-separating triple}), with $X \subseteq V_i \cup \dots V_{i+w}$ and minimum $|X \cap V(G)|$. Equivalently, we want to know whether we can map the vertices of $\Lambda_i^+$ into
sets $X_1, \dots, X_k,\allowbreak X_\text{rest},A,B$ that
satisfy the following conditions:

\begin{itemize}
\item Every vertex belongs to exactly one of the sets $A, X_1,\dots,X_k,X_\text{rest}, B$.
\item $X_1,\dots,X_k,X_\text{rest}$ contain only vertices in $V_{i}\cup \dots \cup V_{i+w}$. 
\item $X_1,\dots,X_k$ contain exactly one vertex each, and that vertex belongs to $G$.   
\item $X_\text{rest}$ contains only dummy vertices and face vertices.
\item $A,B$ contain at least one vertex that belongs to $G$.
\item There is no edge of $\Lambda_i^+$ between $A$ and $B$. 
\end{itemize}

We can express this as a finite-length formula in MSOL, for some fixed $k \leq 7$:

\begin{mdframed}[skipabove=12pt, skipbelow=12pt, nobreak = true]
{
\centerline{MSOL Formula for Finding a Co-Separating Triple}
\bigskip
\noindent$\exists X_1,\dots,X_k,X_\text{rest},A,B \subseteq V(\Lambda_i^+):$ 
\begin{enumerate}
\setlength\itemsep{0.5em}
\item $\forall x \bigvee_{T\in \{X_1,\dots,X_k,X_\text{rest},A,B\}} x\in T$ (`every vertex belongs to at least  one set'), and
\item $\forall x \bigwedge_{T\neq T'\in \{X_1,\dots,X_k,X_\text{rest},A,B\}} x{\in}T {\Rightarrow} x \notin T'$ (`every vertex belongs to at most one set'), and
\item $\forall x \bigwedge_{T \in \{X_1,\dots,X_k,X_\text{rest}\}} x \in T \Rightarrow \neg (V_{i-1}(x) \vee V_{i+w+1}(x))$ (`$X$ is not in outer layers'), and
\item $\bigwedge_{j=1}^k  \exists x: x \in X_j$ (`$X_j$ contains at least one vertex'), and
\item $\forall x,y \bigwedge_{j=1}^k (x\in X_j \wedge y\in X_j \Rightarrow x=y)$ (`$X_j$ contains at most one vertex'), and
\item $\forall x \bigwedge_{j=1}^k (x\in X_j \Rightarrow V_G(x))$ (`each $X_j$ contains only $G$-vertices'), and
\item $\forall x: x\in X_\text{rest} \Rightarrow \neg V_G(x)$ (`set $X_\text{rest}$ contains no $G$-vertices'), and
\item $\bigwedge_{T\in \{A,B\}} \exists x: x\in T \wedge V_G(x)$ (`$A$ and $B$ contain at least one $G$-vertex'), and
\item $\forall x,y: x\in A \wedge y\in B \Rightarrow \neg E(x,y)$ (`no edge between $A$ and $B$')
\end{enumerate}}
\end{mdframed}

Since $\Lambda_i$ is a planar graph with radius at most $w+2$, it has treewidth $O(w)$ \cite{Baker94,eppstein}. Therefore, by \Cref{claim: construct calT+}, $\Lambda_i^+$ has a tree decomposition whose width is in $O(w) \subseteq O(1)$, and we can use Courcelle's theorem to evaluate the above MSOL formula in linear time.

\begin{theorem}[Courcelle's theorem \cite{courcelle1990monadic}]\label{thm: courcelle}
    Every graph property definable in the monadic second-order logic of graphs can be decided in linear time on graphs of bounded treewidth.
\end{theorem}

It should be mentioned here that the use of Courcelle's theorem and MSOL is more powerful than really required. It is not hard, but tedious, to show that one can use bottom-up dynamic programming in the tree decomposition of $\Lambda_i^+$. (The interested reader may refer to \cite{esa_arxiv}, the full-version of \cite{esa_paper}, for an explicit dynamic programming formulation.)

\subsection{Final Algorithm}

The algorithm for testing vertex-connectivity hence proceeds as follows. First pre-process $G$ and duplicate edges to add kite edges where required. Since we never create a face bounded by two edges (bigons), $G$ continues to have at most $4n-8$ edges \cite{Bodendiek1983BemerkungenZE,pach1997graphs}.
Then compute $\Lambda(G)$, the BFS-tree and the layers $V_j$, edge-sets $E_j,E_{j,j{+}1}$, $U_j$ and $L_j$ for all $j \in \{0,\dots,d\}$; all this takes $O(|V(\Lambda(G))|)$ time (\Cref{claim:U and L}).  
For $k=1,\dots,7$, run the following sub-routine to test whether there exists a vertex cut of size $k$; this will necessarily find the minimum such set. 

\begin{mdframed}[skipabove=12pt, skipbelow=12pt, nobreak = true]
{   
\centerline{Sub-Routine for Testing $k$-Vertex Connectivity}

\medskip
\noindent For $i=0,\dots,d$:

\begin{enumerate}
\setlength\itemsep{0.5em}
\item \textbf{Compute $\Lambda_i$:} As all vertices and edge sets are pre-computed, this takes $O(|V(\Lambda_i)|)$ time.

\item 
\textbf{Compute $\calT_i$:} Since $\Lambda_i$ is a planar graph with radius at most $w+2$, it has treewidth $O(w)$, and a corresponding tree decomposition ${\calT_i}$ can be found in $O(w|V(\Lambda_i)|)$ time. We may also assume that ${\calT_i}$ has $O(|V(\Lambda_i)|)$ bags \cite{Baker94,eppstein}.
	
\item \textbf{Compute $\mathcal{T}_i^+$:} By \Cref{claim: construct calT+}, a tree decomposition $\calT_i^+$ of $\Lambda_i^+$ of width $O(width(\calT_i)) \subseteq O(w)$ can be obtained in $O(width(\calT_i)\cdot|\calT_i|) \subseteq O(w|V(\Lambda_i)|)$ time.

\item \textbf{Use Courcelle's theorem:} Since $\calT_i^+$ has a tree decomposition of width $O(w) \subseteq O(1)$, one can test whether the MSOL formula in \Cref{sec: msol} is satisfiable in $O(|V(\Lambda_i^+)|) \subseteq O(|V(\Lambda_i)|)$ time (\Cref{thm: courcelle}). If the MSOL formula returns TRUE, then output $\{X_1, \dots, X_k\}$ as a minimum vertex cut.
\end{enumerate}}
\end{mdframed}
The run-time for each index $i$ is hence $O(w|V(\Lambda_i)|)$. To bound the total run-time, we must bound $\sum_{i=0}^d O(w|V(\Lambda_i)|)$. Since each $\Lambda_i$ uses $w+2$ consecutive layers, any vertex of $\Lambda(G)$ belong to at most $w+2$ graphs of $\{\Lambda_0,\dots,\Lambda_{d}\}$. Therefore, the total run-time is $\sum_{i=0}^{d} O(w|V(\Lambda_i)|) \subseteq O(w^2 |\Lambda(G)|)$. As $w\in O(1)$ and $|\Lambda(G)| \in O(|V(G)|)$, the run-time is linear. This completes the proof of \Cref{thm: linear time}.


\section{Open Problems for Future Work}\label{sec: conclusion}

\begin{figure}
    \centering
    \includegraphics[width=0.75\linewidth]{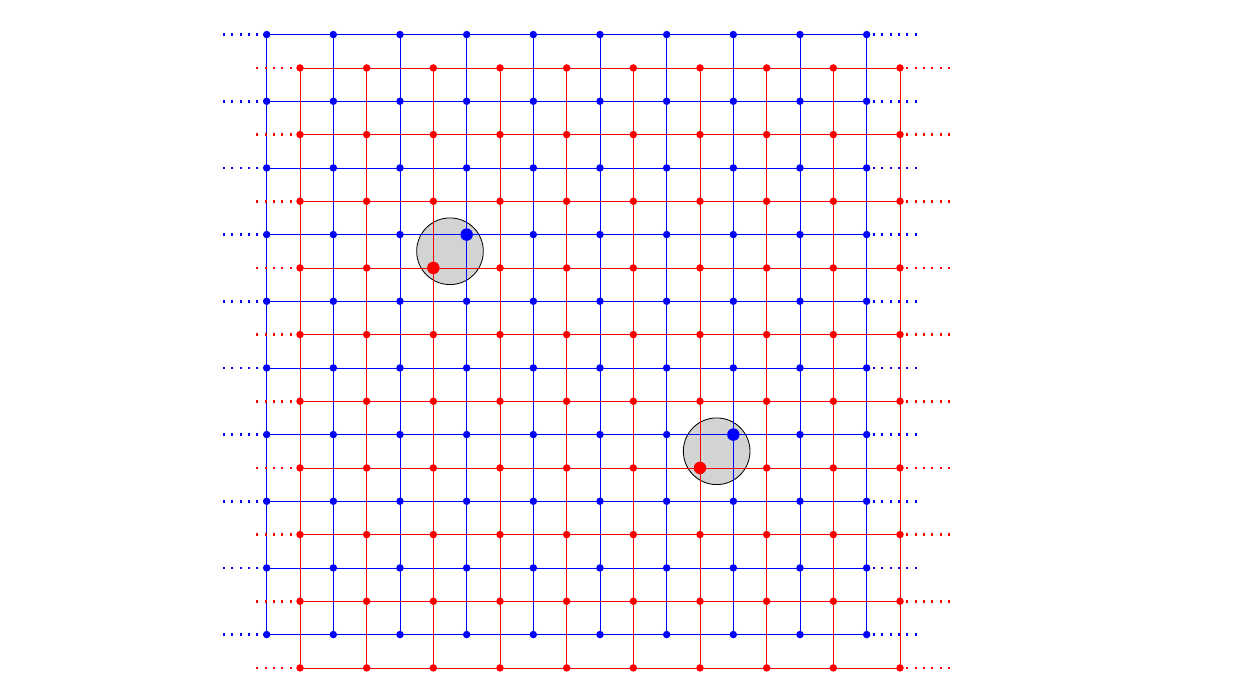}
    \caption{A drawing of a 1-plane graph with $\times$-crossings on a flat cylinder. (The left boundary and right boundary are the same.) The graph is obtained by interleaving two grid graphs, and fusing two pairs of vertices (shown with shaded circles).}
    \label{fig:Xcrossing}
\end{figure}

A key ingredient for the linear-time connectivity testing algorithm for 1-plane graphs without $\times$-crossings is that the vertices of any minimum vertex cut lie within a bounded diameter subgraph of $\Lambda(G)$ (\Cref{thm: co-separating triple and nuclear diameter}). The primary obstacle with $\times$-crossings is that one can construct examples where this property fails to hold. Consider Figure \ref{fig:Xcrossing} which shows two copies of a grid graph interleaved to produce a 1-planar embedding such that each crossing is an \X-crossing. These two graphs can be fused together to form a connected graph $G$ by identifying two pairs of vertices that are far apart from each other. The two fused vertices constitute the only minimum vertex cut of $G$ as each of the two grid graphs is 3-connected. By inserting as many layers in-between as one desires, the distance in $\Lambda(G)$ between the two fused vertices can be increased arbitrarily. In consequence, our techniques to test vertex connectivity cannot be extended to 1-plane graphs with $\times$-crossings. Therefore, our first open question is the following: 

\begin{open_problem}
\label{open:1}
Can the vertex connectivity of all 1-plane graphs be computed in linear time?
\end{open_problem}

Throughout the paper, we assumed that the input graph comes with a 1-planar embedding. We did this since testing 1-planarity is NP-hard. We do not know whether it might be easier to test if a graph has a 1-planar embedding without $\times$-crossings.

\begin{open_problem}
Is it NP-hard to test whether a given graph has a 1-planar embedding without $\times$-crossings?
\end{open_problem}









\bibliography{References}

\end{document}